%% file: main.tex
\documentclass{article}
\usepackage{iclr2027_conference,times}

\input{math_commands.tex}

\usepackage[utf8]{inputenc}
\usepackage[T1]{fontenc}
\usepackage{hyperref}
\usepackage{url}
\usepackage{booktabs}
\usepackage{amsfonts}
\usepackage{amsmath}
\usepackage{amssymb}
\usepackage{amsthm}
\usepackage{nicefrac}
\usepackage{microtype}
\usepackage{xcolor}
\usepackage{graphicx}
\usepackage{tabularx}
\usepackage{array}
\newcolumntype{P}[1]{>{\raggedright\arraybackslash}p{#1}}
\usepackage{tikz}
\usepackage{capt-of}
\usepackage{latexml}
\usepackage{etoolbox}
\AtBeginEnvironment{table}{\setlength{\abovecaptionskip}{0pt}\setlength{\belowcaptionskip}{6pt}}
\usetikzlibrary{arrows.meta,positioning,fit}

\newtheorem{proposition}{Proposition}

\newcommand{\AuthSets}{\mathrm{AuthSets}}
\newcommand{\bflat}{\textsc{B-Flat}}
\newcommand{\bgraph}{\textsc{B-Graph}}
\newcommand{\blore}{\textsc{B-Native}}

\title{Audience-Bound Persistent Memory:\\ Authorization Across the Memory Lifecycle}

\author{Sibo Liu\\Independent Researcher\\\texttt{abc-paper@outlook.com}}

\iclrfinalcopy
\makeatletter
\patchcmd{\@maketitle}{Published as a conference paper at ICLR 2027}{Preprint}{}{\PackageError{arxiv-preprint}{Could not set preprint header}{}}
\makeatother
\hypersetup{hidelinks,pdftitle={Audience-Bound Persistent Memory: Authorization Across the Memory Lifecycle},pdfauthor={Sibo Liu},pdfsubject={Audience-bound persistent memory}}

\begin{document}
\raggedbottom %
\setlength{\textfloatsep}{12pt plus 2pt minus 2pt}\setlength{\floatsep}{10pt plus 2pt minus 2pt}\setlength{\intextsep}{10pt plus 2pt minus 2pt}
\input{generated/h3_macros}\input{generated/h3_arms}\input{generated/h3_sign}\input{generated/h1_exposure}\input{generated/dose_macros}\input{generated/topk_macros}

\maketitle

\begin{abstract}
\unskip
A personal language agent that acts for its owner across private and shared conversations can
learn a fact from one audience and later place it in the context it assembles for another. We study
\emph{authorization before context} across the whole memory lifecycle. Each memory item carries the
audience present when it was recorded; derived items are partitioned by audience, receive the
intersection of their sources' audiences, or are suppressed; an audience widens only by an explicit,
object-specific grant; and an item enters a model attempt only when every current viewer belongs to
one of its authorized audiences, with unresolved viewers failing closed to public-only. Under
explicit identity, provenance and complete-mediation assumptions, this admission is sound and
policy-complete on the exact assembled context, enforced by exclusion rather than by model behavior.
We realize it in two independently persisted reference architectures, a flat store and a
relationship graph, and, descriptively, in a native agent-memory runtime. In a prospectively frozen
confirmation over 10{,}000 multi-party histories,  no forbidden item entered any
architecture's context, whereas unscoped retrieval exposed forbidden items in \HoneUnscopedPct{} of
its contexts. Entitled recall matched policy-equivalent baselines exactly and exceeded unscoped
retrieval by 0.30 Recall@5, with a Holm-confirmed advantage that grows with distractors. No
architecture produced a wrong-principal substitution, but unscoped substitutions were too rare to
establish the prespecified joint decision.
\end{abstract}

\input{sections/01_introduction}

\input{sections/02_related}
\input{sections/03_model}
\input{sections/04_architecture}
\input{sections/05_evaluation}
\input{sections/06_results}
\input{sections/07_limitations}

\input{sections/08_conclusion}

\input{sections/09_statements}

\bibliography{references}
\bibliographystyle{iclr2027_conference}

\appendix
\input{sections/A_formal}
\input{sections/B_protocol}
\input{sections/C_accounting}

\input{sections/D_related_matrix}
\input{sections/E_details}

\end{document}

%% file: math_commands.tex
\usepackage{amsmath,amsfonts,bm}

\def\eqref#1{equation~\ref{#1}}

\def\1{\bm{1}}

\DeclareMathAlphabet{\mathsfit}{\encodingdefault}{\sfdefault}{m}{sl}
\SetMathAlphabet{\mathsfit}{bold}{\encodingdefault}{\sfdefault}{bx}{n}

%% file: generated/h3_macros.tex
\newcommand{\HJointOrig}{not established}
\newcommand{\HMissOrig}{85}
\newcommand{\HSubFlatLoOrig}{$-0.0028$}
\newcommand{\HSubFlatHiOrig}{$+0.0023$}
\newcommand{\HSubFlatPOrig}{$1$}
\newcommand{\HSubGraphLoOrig}{$-0.0027$}
\newcommand{\HSubGraphHiOrig}{$+0.0023$}
\newcommand{\HSubGraphPOrig}{$1$}
\newcommand{\HSubRejOrig}{not rejected}
\newcommand{\HSlopeFlatOrig}{$+0.130$}
\newcommand{\HSlopeGraphOrig}{$+0.130$}
\newcommand{\HSlopePOrig}{$<10^{-6}$}
\newcommand{\HSlopeRejOrig}{rejected}
\newcommand{\HRecallPOrig}{$<10^{-6}$}
\newcommand{\HRecallEstOrig}{$+0.298$}
\newcommand{\HRecallRejOrig}{rejected}
\newcommand{\HJudgedOrig}{29{,}915}
\newcommand{\HPendingOrig}{0}
\newcommand{\HMissingStateOrig}{85}
\newcommand{\HJointAmend}{not established}
\newcommand{\HMissAmend}{18}
\newcommand{\HSubFlatLoAmend}{$+0.0000$}
\newcommand{\HSubFlatHiAmend}{$+0.0010$}
\newcommand{\HSubFlatPAmend}{$1$}
\newcommand{\HSubGraphLoAmend}{$-0.0001$}
\newcommand{\HSubGraphHiAmend}{$+0.0010$}
\newcommand{\HSubGraphPAmend}{$1$}
\newcommand{\HSubRejAmend}{not rejected}
\newcommand{\HSlopeFlatAmend}{$+0.130$}
\newcommand{\HSlopeGraphAmend}{$+0.130$}
\newcommand{\HSlopePAmend}{$<10^{-6}$}
\newcommand{\HSlopeRejAmend}{rejected}
\newcommand{\HRecallPAmend}{$<10^{-6}$}
\newcommand{\HRecallEstAmend}{$+0.298$}
\newcommand{\HRecallRejAmend}{rejected}
\newcommand{\HJudgedAmend}{29{,}982}
\newcommand{\HPendingAmend}{0}
\newcommand{\HMissingStateAmend}{18}
\newcommand{\HJointSummary}{not established in either view}

%% file: generated/h3_arms.tex
\newcommand{\HSubUOrig}{7}
\newcommand{\HJudUOrig}{9{,}984}
\newcommand{\HSubFOrig}{0}
\newcommand{\HJudFOrig}{9{,}965}
\newcommand{\HSubGOrig}{0}
\newcommand{\HJudGOrig}{9{,}966}
\newcommand{\HSubUAmend}{7}
\newcommand{\HJudUAmend}{9{,}997}
\newcommand{\HSubFAmend}{0}
\newcommand{\HJudFAmend}{9{,}993}
\newcommand{\HSubGAmend}{0}
\newcommand{\HJudGAmend}{9{,}992}

%% file: generated/h3_sign.tex
\newcommand{\HSignFor}{7}
\newcommand{\HSignAgainst}{0}
\newcommand{\HSignP}{0.0078}

%% file: generated/h1_exposure.tex
\newcommand{\HoneCells}{480{,}000}
\newcommand{\HoneTargetCells}{200{,}000}
\newcommand{\HoneOtherCells}{180{,}000}
\newcommand{\HoneUnscopedCells}{100{,}000}
\newcommand{\HoneUnscopedItems}{220{,}385}
\newcommand{\HoneUnscopedExposed}{82{,}266}
\newcommand{\HoneUnscopedPct}{82\%}

%% file: generated/dose_macros.tex
\newcommand{\DoseArch}{0.974}
\newcommand{\DoseUnscZero}{0.974}
\newcommand{\DoseUnscEight}{0.616}
\newcommand{\DoseAdv}{0, +0.056, +0.265, +0.358, +0.358}
\newcommand{\DoseAdvMax}{0.358}
\newcommand{\DoseHalfWidth}{0.008}
\newcommand{\AbsArch}{0.972}
\newcommand{\AbsUnsc}{0.674}
\newcommand{\AbsSilo}{0.740}

%% file: generated/topk_macros.tex
\newcommand{\TopkArchFull}{10}
\newcommand{\TopkUnscFull}{18}
\newcommand{\TopkSiloMax}{0.750}
\newcommand{\TopkArchOne}{0.580}
\newcommand{\TopkUnscOne}{0.527}

%% file: sections/01_introduction.tex
\section{Introduction}
\label{sec:intro}
\unskip

A personal language agent acts as a delegated proxy for its owner. On the owner's
behalf it takes part in private conversations with the owner and in one-to-one and
group conversations with other participants, who may be people or other agents,
across transports such as email, team chat and text messages. What it learns in all of these conversations accumulates in one long-term memory,
although each fact was shared with a particular \emph{audience}: the participants
present when it was said. For each
model call, the agent assembles a context from the current conversation's recent
messages and from long-term memory, which holds both what was said and what is derived from it, such as summaries
and knowledge-graph facts. The resulting threat is compositional: a fact shared
with one audience could be stored,
transformed, retrieved on a later turn and placed in the context assembled for a
different audience. Once it is there, later defenses are too late. The model has
already read it; redacting the response cannot undo that exposure, and scoring
the answer cannot detect it, because a model can answer acceptably even when its
context (instructions, conversation history and retrieved memory) contains a
forbidden fact.

\begin{figure}[b]
\centering
\iflatexml
\includegraphics[width=397.48499pt]{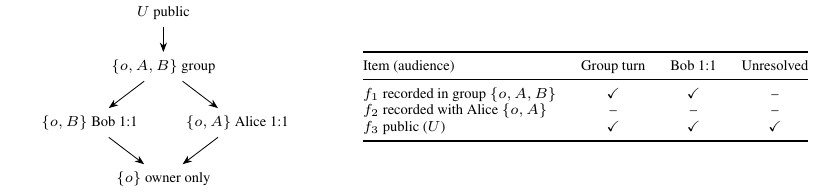}
\else
\begin{minipage}[c]{0.40\linewidth}
\centering
\includegraphics[width=125.73991pt,height=92.98267pt]{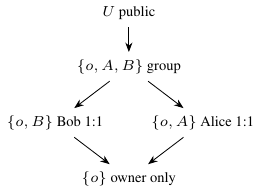}
\end{minipage}\hfill
\begin{minipage}[c]{0.58\linewidth}
\centering
\scriptsize
\begin{tabular}{@{}lccc@{}}
\toprule
Item (audience) & Group turn & Bob 1:1 & Unresolved \\
\midrule
$f_1$ recorded in group $\{o,A,B\}$ & \checkmark & \checkmark & -- \\
$f_2$ recorded with Alice $\{o,A\}$ & -- & -- & -- \\
$f_3$ public ($U$) & \checkmark & \checkmark & \checkmark \\
\bottomrule
\end{tabular}
\end{minipage}
\fi
\caption{Admission by audience membership. Left: the audience lattice; an item flows downward, to containers whose viewers
are a subset of its audience, never upward. Right: which items enter each turn's
context. Bob's one-to-one recalls the group fact $f_1$ but never Alice's fact
$f_2$; a turn with unresolved viewers fails closed to public items.}
\label{fig:example}
\end{figure}

Three common designs fall short. Relevance-only retrieval ignores who is present
on the current turn. Output filtering acts after the model has read the context.
Per-user or per-session silos avoid exposure by giving up legitimate recall: a
fact shared in a group that includes Bob is exactly what Bob's later one-to-one
conversation is entitled to recall (Figure~\ref{fig:example}). We instead treat
the problem as one of authorization in the memory system and the agent harness
around the model, not as a behavior the model should learn. We keep one shared
memory but label each item with its audience, read from the transport when the
item is written (always including the owner), carry the label through every
transformation, and check every path from memory into a model's context before
the model is called: an item is admitted only when every current viewer already belongs to one of its
authorized audiences.

This is a relationship check in the style of \citet{pang2019zanzibar}; related
predicates appear in access control, in authorization-first retrieval
over static documents \citep{afr2026}, and at a personal agent's
memory-to-context transition (\S\ref{sec:related}). What is missing is its \emph{lifecycle}: how
labels are bound at write time, computed for derived memory, widened only by
explicit decisions, kept current, and enforced for every physical model call
across different stores. Because a fully mediated retriever never returns an
inadmissible item, the central empirical question is the cost in recall: whether
evidence the audience is entitled to see is still found, and how retrieval behaves
when highly relevant facts about \emph{other} participants compete for the same
slots.

\paragraph{Contributions.}
\begin{enumerate}
  \item \textbf{Lifecycle authorization model} (\S\ref{sec:model}): origin-audience
  binding at write time; four closed derivation dispositions; exact,
  non-unioned declassification grants that never carry over to derived items;
  and admission against one authorization snapshot per physical model attempt.
  \item \textbf{Conditional guarantee} (\S\ref{sec:guarantee}): under the identity, provenance and complete-mediation assumptions of \S\ref{sec:tcb}, no model call's
  context contains an inadmissible item (soundness), and an exhaustive query
  returns every admissible one (policy completeness). In both reference
  architectures an excluded item also never occupies a candidate slot, so it cannot
  displace an admissible one.
  \item \textbf{Three realizations} (\S\ref{sec:arch}): two independently
  persisted reference architectures that share only the policy package, and a realization inside an existing agent-memory runtime, validated descriptively.
  \item \textbf{A prospectively frozen evaluation} (\S\ref{sec:eval}): a
  10{,}000-history paired confirmation with a prespecified non-inferiority margin,
  closed testing, Holm-controlled superiority, conformance testing and complete
outcome accounting.
   No forbidden item entered any of \HoneTargetCells{} exact contexts; all eight
non-inferiority checks pass with no recall cost against
policy-equivalent baselines, and recall superiority and the dose slope are
confirmed; substitutions (0 vs.\ 7) were too rare to confirm H3a/b, so the joint
decision is not established (\S\ref{sec:results}).{}
  \unskip
\end{enumerate}

%% file: sections/02_related.tex
\section{Related Work}
\label{sec:related}

\paragraph{Authorization before context.} Authorization-First Retrieval
\citep{afr2026} argues that, in multi-agent retrieval over documents,
authorization must precede any learned component, since retrieve-then-filter
pipelines expose unauthorized content structurally. Filtered vector search shows
that attribute filters break the recall assumptions of approximate indexes
\citep{chronis2025filtered}. \citet{liu2026authorization} applied an
audience-membership test at a personal agent's memory-to-context transition,
proved one-way confinement and fail-closed behavior for the rule, and reported
no forbidden inclusions on a 79-scenario synthetic suite against unscoped
projections that include them by construction. We add what they leave open: derived labels, exact grants, lifecycle currentness, per-attempt admission, independent realizations and a prospectively powered recall study.

\paragraph{Governed and multi-user memory.} Collaborative Memory
\citep{rezazadeh2025collaborativememory} shares memory among users and agents
through configured, time-varying access graphs and provenance-tagged fragments;
governed shared-memory services use scopes and roles
\citep{memclaw2026,superlocalmemory2026}; and multi-user agents and
deployment-time memorization expose the privacy failure surface
\citep{yang2026multiuser,chen2026deploymentmemorization}. Our audiences are derived from the transport identities of the participants present rather than configured, and the endpoint is the exact context of each model attempt.

\paragraph{Privacy and memory benchmarks.} Contextual-integrity and privacy
benchmarks for memory and multi-user agents mostly score model output or task
utility \citep{mireshghallah2025cimemories,pisas2026,gatemem2026}, and
multi-party memory benchmarks score utility
\citep{yang2026groupmembench,evermembench2026}. We score the assembled context itself (GateMem's construct limitation is in
App.~\ref{app:related}).

\paragraph{Authority, integrity and information flow.} Other work asks what
memory authorizes an agent to \emph{do} or whether memory can be trusted
\citep{authmembench2026,eal2026,ppmf2026,tmanm2026,mapgraph2026,memsecbench2026,mutmem2026};
we share its write-time origin binding but address read-audience confidentiality
only. Our labels and grants follow lattice and decentralized information-flow
control and declassification
\citep{bell1976secure,denning1976lattice,myers1997decentralized,sabelfeld2009declassification},
and our derived-data rules follow provenance-based policy
\citep{bates2013provenance,denhartog2016datafusion}; agent information-flow
systems \citep{costa2025fides,debenedetti2025camel,palumbo2026forge} instead
govern control and effect flows under prompt injection. We implement only the
recipient slice of contextual integrity \citep{nissenbaum2009privacy}.
Appendix~\ref{app:related} gives a detailed comparison.

%% file: sections/03_model.tex
\section{Lifecycle Authorization Model}
\label{sec:model}

This section extends a single audience-membership admission rule to writes,
derivations, grants and model attempts.

\subsection{Participants, viewers and audiences}
\label{sec:labels}

Let $U$ be the set of \emph{participants}, people or agents, with a distinguished
owner $o\in U$ and an anonymous participant $\star\in U$ that stands for any viewer
who cannot be resolved. A \emph{container} $C$ is one conversation turn; its
\emph{viewers} $V(C)\subseteq U$ are the participants present on that turn. The agent takes part on the owner's behalf, so $o\in V(C)$ always. When the
viewer evidence (the transport identities of those present, \S\ref{sec:tcb}) is
missing, stale, low-confidence or conflicting, $V(C)=U$ (fail closed). Groups and
channels are containers, not participants, and an audience is identified by its
member set, not by the channel on which it is observed.

Each memory item $m$ carries an \emph{audience} $\mathrm{Aud}(m)\subseteq U$, the
participants who may see it; the owner belongs to every audience. An audience has
one of three kinds: \emph{public}, $\mathrm{Aud}(m)=U$, which any participant may
see; \emph{owner-private}, $\{o\}$; and \emph{shared}, the resolved viewers $A$ of
the container where $m$ was recorded, with $o\in A\subsetneq U$. Only a public
audience contains $\star$. We use \emph{participant} throughout; the frozen
protocol's names \emph{principal postings} and \emph{wrong-principal substitution}
use \emph{principal} in the same sense.

The lifecycle adds three objects. \emph{Write-time binding} sets
$\mathrm{Aud}(m)$ of a recorded item to the resolved viewers of its origin
container (or $U$ for a public write) and fails closed: a write whose origin viewers cannot be determined is refused rather than bound
to a guessed audience. A \emph{derived} item receives an
audience from its sources (\S\ref{sec:derivation}). A \emph{grant} adds one exact,
object-specific audience of resolved participants; $\mathrm{Grants}(m,t)$ holds
the grants in force at time $t$, and the authorized audiences
$\AuthSets(m,t)=\{\mathrm{Aud}(m)\}\cup\mathrm{Grants}(m,t)$ are evaluated
separately, never unioned. Otherwise an audience can only be narrowed
(App.~\ref{app:revocation}).

\subsection{Admission}
\label{sec:admission}

Let $\mathrm{Current}(m,t)$ hold when $m$ and every transitive source are
provenance-complete at time $t$ and none is superseded, frozen, revoked or
quarantined. Container $C$ at time
$t$ admits
\begin{equation}
\label{eq:policy}
\mathrm{Adm}(C,t)=\{\,m:\ \mathrm{Current}(m,t)\ \wedge\
\exists A\in\AuthSets(m,t):\ V(C)\subseteq A\,\}.
\end{equation}
Without grants and lifecycle state this is exactly the rule of
\citet{liu2026authorization}. Table~\ref{tab:states} instantiates it; admission
is by exclusion, so an inadmissible item never reaches the model.

\begin{table}[t]
\centering
\small
\caption{Eq.~(\ref{eq:policy}) by viewer evidence (every admitted item must also
be current).}
\label{tab:states}
\begin{tabular}{@{}P{0.36\linewidth}P{0.58\linewidth}@{}}
\toprule
Viewer evidence for $C$ & Admitted memory \\
\midrule
Owner only, $V(C)=\{o\}$ & Every item; reading widens no audience \\
Resolved viewers & Items with some $A\in\AuthSets(m,t)$, $V(C)\subseteq A$ \\
Missing, stale, low-confidence or conflicting & $V(C)=U$: public items only \\
\bottomrule
\end{tabular}
\vspace{0.7em}
\caption{Derivation dispositions for an item derived from sources $S$.}
\label{tab:derivation}
\begin{tabular}{@{}lP{0.36\linewidth}P{0.38\linewidth}@{}}
\toprule
Disposition & Resulting audience & When \\
\midrule
Partition & One item per source audience & Sources may stay separate \\
Intersection & $\bigcap_{s\in S}\mathrm{Aud}(s)$ (always $\ni o$) & Indivisible artifact \\
Declassification & Adds one grant; $\mathrm{Aud}(m)$ unchanged & Authenticated, object-specific decision \\
Suppress / quarantine & None: not created, or quarantined and never current & Incomplete provenance, or a source reachable only via a grant \\
\bottomrule
\end{tabular}
\end{table}

\subsection{Derivation and grants}
\label{sec:derivation}

Every persisted transformation takes one of four closed dispositions
(Table~\ref{tab:derivation}). Derivation reads each source's $\mathrm{Aud}$,
never its grants, so a grant cannot become authority for a new item that might
outlive it.

\subsection{Threat model, attempts and trusted computing base}
\label{sec:attempt}
\label{sec:tcb}

For a container $C$, the protected assets are the items, derived artifacts and
fields that are not admissible for $C$ under Eq.~(\ref{eq:policy}). Table~\ref{tab:threats} maps the
failures we consider to the mechanism that addresses each. The endpoint is the
exact context of each \emph{physical} model attempt, checked against one
authorization snapshot pinned at its linearization point and sealed before the
model is called; later mutations govern later attempts, and retries or fallbacks
are new attempts. The \emph{oracle} is the code that evaluates
Eq.~(\ref{eq:policy}); Figure~\ref{fig:lifecycle}(a) shows where it acts.

\paragraph{Transports and identity.} A \emph{transport} is the messaging service
that carries a conversation, such as an email service, a team-chat workspace or a
text-messaging network. Each names senders and recipients by its own
authenticated identifier: an email address, a workspace account identifier, a
phone number. An \emph{identity resolver} maps each one to a single participant, and ambiguous or
conflicting resolution fails closed. The trusted computing base is therefore these
transport identities, the resolver, provenance capture, the oracle, stored audiences
and grants, and every read path's call to the oracle; the model, rankers, answers,
output filters and stored text are untrusted.

\begin{table}[t]
\centering
\small
\caption{Threats and the mechanism addressing each. Forged transport identities and
wrong identity resolution are trusted-base failures, not defended.}
\label{tab:threats}
\begin{tabular}{@{}P{0.56\linewidth}P{0.38\linewidth}@{}}
\toprule
Threat or failure & Addressed by \\
\midrule
A viewer elicits a fact recorded for another audience & Admission by exclusion (P1) \\
Relevant facts about another participant compete for slots & Authorization before ranking; slot noninterference (\S\ref{sec:guarantee}, H3) \\
Missing, stale, low-confidence or conflicting viewer evidence & Fail closed: public-only reads (P3), refused writes \\
Poisoned memory tries to widen its own audience & Write-time binding (P4) \\
Summaries, merges and derivations mix audiences & Derivation dispositions (P5) \\
Direct lookup, graph or aggregate paths bypass retrieval & Complete mediation (Prop.~\ref{prop:guarantee}) \\
Source or grant changes during an attempt & One snapshot per attempt \\
\bottomrule
\end{tabular}
\vspace{0.7em}
\caption{Properties of Eq.~(\ref{eq:policy}). P1--P4 extend the single-rule
properties; P5 is new.}
\label{tab:props}
\begin{tabular}{@{}lP{0.68\linewidth}@{}}
\toprule
Property & Statement \\
\midrule
P1 one-way confinement & $m\in\mathrm{Adm}(C,t)\Rightarrow V(C)\subseteq A$ for some $A\in\AuthSets(m,t)$ \\
P2 anti-monotonicity & $V(C_1)\subseteq V(C_2)\Rightarrow\mathrm{Adm}(C_2,t)\subseteq\mathrm{Adm}(C_1,t)$ \\
P3 fail-closed soundness & $V(C)=U\Rightarrow$ only current public items are admitted \\
P4 poisoning containment & An injected item's audience is fixed at write time and cannot widen itself \\
P5 derivation non-expansion & An intersection item's audience lies within every source's; no grant is inherited \\
\bottomrule
\end{tabular}
\end{table}

\subsection{Guarantee}
\label{sec:guarantee}

The four properties of the single-rule setting \citep{liu2026authorization} carry
over with $\AuthSets$ in place of $\mathrm{Aud}$, and derivation adds a fifth
(Table~\ref{tab:props}; proofs in App.~\ref{app:formal}).

Let $\widehat{\mathrm{Adm}}(C,t)$ be what the \emph{implementation} admits and
$Q(C)$ the items satisfying the query's non-authorization filters.

\begin{proposition}[Conditional soundness and policy completeness]
\label{prop:guarantee}
If the trusted computing base is correct and every memory-to-context path invokes
the oracle on one snapshot, then (i)
$\widehat{\mathrm{Adm}}(C,t)\subseteq\mathrm{Adm}(C,t)$, and (ii) a backend that
declares exhaustive retrieval returns every item of $\mathrm{Adm}(C,t)\cap Q(C)$.
\end{proposition}

Part (i) says the implementation admits nothing outside Eq.~(\ref{eq:policy});
part (ii) says it rejects nothing inside it. Neither is a retrieval-quality
promise, since ranking and a finite $k$ can still lose entitled evidence, so
recall is measured.

\paragraph{Context exclusion and slot noninterference.} The proposition guarantees
\emph{context exclusion}: the model never sees an inadmissible item. Both reference
architectures add \emph{slot noninterference}: every candidate is authorized before
ranking by per-item scores fixed before authorization, so an inadmissible item never occupies a
fetch or context slot and cannot displace an admissible one (confirmed in all 10{,}000
histories, \S\ref{sec:results-wrong-principal}). Full \emph{selection
noninterference} would also remove inadmissible text from shared ranking statistics,
which requires per-audience index statistics; it is not part of the guarantee.

\paragraph{Scope.} Grant revocation, deletion and member removal are simple under
Eq.~(\ref{eq:policy}) and take effect from the next attempt's snapshot
(App.~\ref{app:revocation}). Out of scope are transport authentication, same-audience contextual-integrity
norms, physical erasure and unlearning, revocation within a sealed attempt, side
channels, action authorization and general prompt-injection defense.

%% file: sections/04_architecture.tex
\section{Architecture and Implementation}
\label{sec:arch}

\begin{figure}[t]
\centering
\includegraphics[width=385.18538pt,height=256.00645pt]{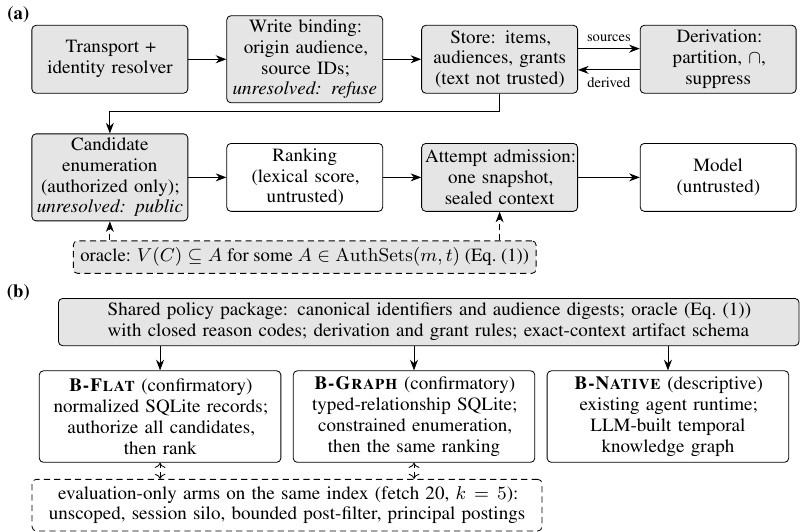}
\caption{(a) Write--transform--read lifecycle. Shaded components belong to the
trusted computing base, and italics mark the fail-closed rules; the same oracle constrains candidate enumeration before
ranking and re-checks every physical model attempt; ranking, stored text and
the model are not trusted to enforce confidentiality. (b) Realizations: two
independently persisted reference architectures share only the policy package
and are compared with evaluation-only arms; the native realization is reported
descriptively.}
\label{fig:lifecycle}\label{fig:arch}
\end{figure}

\subsection{Common contract and reference architectures}
\label{sec:arms}

Figure~\ref{fig:arch}(b) shows the realizations. All share one code-owned policy
package: canonical identifiers and audience digests, the oracle for
Eq.~(\ref{eq:policy}) and Table~\ref{tab:states} with closed reason codes, the
derivation and grant rules of \S\ref{sec:derivation}, and the exact-context
artifact schema. They share neither storage nor candidate queries, so differential conformance
testing between them is not tautological, and each constrains candidates \emph{before} ranking. \bflat{} is a durable normalized store with
separate item, audience, source, derivation and grant records; it authorizes the
exhaustive candidate set and then ranks deterministically. \bgraph{} is a durable
relationship store whose typed participant, audience, item, provenance, derivation
and grant relations constrain candidate enumeration before the same ranking and
context assembly. Both persist to SQLite through independent schemas and query
paths and are the two confirmatory architecture classes
\unskip.

Four evaluation-only arms share the same queries, memory state, model, prompts and deterministic lexical index (built once; fetch budget 20,
$k=5$): unscoped relevance retrieval, a fixed session silo, a bounded post-filter (rank the top 20 without authorization, then drop inadmissible
ones before assembly; not output filtering), and principal postings. The last stores, for each participant $p$, the postings
$N_p=\{(m,A):p\in A\in\AuthSets(m,t)\}$ and admits $m$ when some pair $(m,A)$
appears in the postings of every current viewer; it is policy-equivalent to the
oracle, so it is a strong baseline.

\subsection{Native realization}

\blore{} realizes the same contract inside an existing personal-agent memory
runtime whose long-term store includes a temporal knowledge graph built by an LLM
extraction pipeline. We extended the existing component responsible for each write, derivation,
retrieval, context-carrier and provider boundary rather than adding a parallel
authorization service; native-specific mechanisms are in App.~\ref{app:native-mech}. It is validated descriptively, outside the confirmatory family
(\S\ref{sec:results-native}), for persistence, restart, concurrency and mediation in
the research configuration\unskip.

%% file: sections/05_evaluation.tex
\section{Evaluation Design}
\label{sec:eval}

\begin{figure}[t]
\centering
\includegraphics[width=352.94612pt,height=118.14383pt]{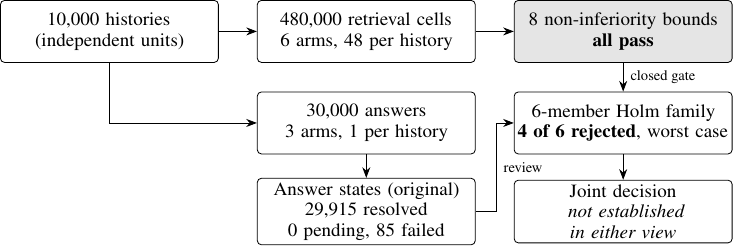}
\caption{Confirmatory pipeline and status. Retrieval gates are complete and pass;
 every contract-valid answer is judged, and the worst-case supplement decides the Holm
family (\S\ref{sec:results-wrong-principal}).{}  Answer states include all saved judgments;{} the amended recovery view differs only in answer states
(App.~\ref{app:accounting}).}
\label{fig:flow}
\end{figure}

Protocol, sample plan and decision rules were frozen before any confirmatory outcome was inspected
(amendments in App.~\ref{app:deviations}); Table~\ref{tab:hyp} and Figure~\ref{fig:flow} summarize them.

\begin{table}[!ht]
\centering
\small
\caption{Hypotheses, tests and outcomes.{} H1 holds on every exact context; H2 and the Holm family
(decided under the worst-case supplement) are complete.}
\label{tab:hyp}
\setlength{\tabcolsep}{4pt}
\begin{tabular}{@{}llll@{}}
\toprule
& Claim & Test & Status \\
\midrule
H1 & No forbidden item in any context & Exposure per attempt; suites & \textbf{0 of 200{,}000} (\S\ref{sec:results-conformance}) \\
H2 & Entitled recall is non-inferior & 8 paired bounds $>-0.05$ & \textbf{All 8 pass} (\S\ref{sec:results-recall}) \\
H3a/b & Fewer wrong-principal substitutions & Answer-level rate; Holm &  Null not rejected (Tab.~\ref{tab:h3}){} \\
H3c & Advantage grows with distractors & Slope on $\log_2(d{+}1)$; Holm &  Null rejected (Tab.~\ref{tab:h3}){} \\
H4a/c/d & Derivation never widens audiences & Transformation, mutation & Cases pass (\S\ref{sec:results-derivation}) \\
H4b & Intersection keeps availability & Intersection vs.\ suppression & 8/8 vs.\ 0/8 (App.~\ref{app:supporting-full}) \\
\bottomrule
\end{tabular}
\end{table}

\subsection{Design}
\label{sec:eval-data}

\paragraph{Confirmation frame.} The frame is 10{,}000 independently generated
multi-party histories (draws 40--10{,}039, 1{,}250 in each of eight themes), from a
controlled event law whose ledger fixes participants, containers, audiences,
viewers, gold evidence and answer targets before language rendering
\unskip. Histories cover
cross-channel recall, same-predicate distractors about \emph{other} participants
at doses $d\in\{0,1,2,4,8\}$ on the temporal-update question (paired advantage
$\delta(d)$; the six-question endpoint uses $d=8$), and abstention when only
unauthorized evidence exists. Draws 0--39, including the 32-history pilot, are
excluded. Inference concerns this generator, not natural prevalence.

\paragraph{Arms, models and other evidence.} Each history yields 48 paired
retrieval cells over the six arms of \S\ref{sec:arms} and one answer each from
unscoped retrieval, \bflat{} and \bgraph{}. All confirmation model calls use one pinned model
(\texttt{gpt-5.6-luna}, medium reasoning, no fallback, observed identity checked; limits in App.~\ref{app:protocol}), and one shared lexical index per history (FTS5 BM25) ranks candidates without embeddings. An external frame of independently authored corpora with their original gold labels (500 LongMemEval \citep{wu2025longmemeval}, 1{,}981 LoCoMo \citep{maharana2024locomo} and 68 EverMemBench \citep{evermembench2026} questions) shows recall parity across all six arms (App.~\ref{app:external}).

\subsection{Inference}
\label{sec:stats}

For history $h$ and an architecture--comparator pair, $D_h\in[-1,1]$ is the mean
over six equally weighted questions of the paired authorized Recall@5 difference;
the estimand is $\mu=\mathbb{E}[D_h]$, with no imputation. Because histories are
independent and bounded, the frozen test is the one-sided empirical-Bernstein
bound of \citet{maurer2009empirical},
$\bar D-\sqrt{2V_n\ln(2/\alpha)/n}-7r\ln(2/\alpha)/(3(n-1))$, with $n=10{,}000$,
sample variance $V_n$ and statistic range $r$ ($r=2$ for $D_h$). H2 passes only if
all eight bounds at $\alpha=0.025$ exceed $-0.05$, i.e., on average at most one extra complete miss of gold evidence per twenty
questions. The gate then opens a
six-member Holm family at $\alpha=0.05$ (recall superiority, reduced substitution
and the H3c slope for each architecture), with p-values from the frozen tests
(Table~\ref{tab:protocol}). The
plan gave a minimum simultaneous Monte Carlo lower bound of 0.935 on joint power
over 40 scenarios (target 0.90).

\paragraph{Answer states and views.} Each answer is resolved, pending semantic review, a terminal
failure or an unresolved recovery (App.~\ref{app:accounting}). We report the original protocol
and an amended view in which terminal citation-contract failures had one recorded
recovery attempt, and select neither as primary
\unskip.

\paragraph{Answer judging.} Under an approved amendment, AI judges apply the frozen rubric to
non-exact answers. Each judging route is a recorded provenance stratum, no judge saw keys or prior
labels, and every judgment is schema- and citation-validated before it is saved (App.~\ref{app:judging}). \unskip

%% file: sections/06_results.tex
\section{Results}
\label{sec:results}

{}

\subsection{No forbidden item reaches the model (H1)}
\label{sec:results-conformance}

 Exclusion holds by construction (Proposition~\ref{prop:guarantee}), and the
campaign confirms it: none of the \HoneTargetCells{} exact contexts assembled by
\bflat{} and \bgraph{}, nor the \HoneOtherCells{} of the silo, post-filter and postings arms, contained a
forbidden item, whereas unscoped retrieval placed \HoneUnscopedItems{} forbidden items
into \HoneUnscopedExposed{} of its \HoneUnscopedCells{} contexts (\HoneUnscopedPct), per the
scorer's stored exact-context counts.{} Conformance testing agrees: 94 differential cases passed, 71 mutants were killed and
24 transformation cases passed on an earlier revision, and all 30 selected checks
passed on the evaluated one \unskip.

\subsection{Entitled recall is non-inferior (H2)}
\label{sec:results-recall}

Authorization before ranking costs no recall. All eight prespecified non-inferiority
checks pass at the $-0.05$ Recall@5 margin (Table~\ref{tab:ni}), identically for both
architectures and answer views. Against the policy-equivalent
bounded post-filter and principal postings, the paired difference is exactly zero in every history (eight distractors never exhausted the post-filter's 20-item window); against unscoped and session-silo retrieval, the architectures recover more entitled evidence ($+0.298$ and $+0.233$); the ordering holds at every $k$ from 1 to 20 (Figure~\ref{fig:topk}).

\iflatexml
\begin{table}[t]\centering
\caption{Entitled-recall non-inferiority (10{,}000 histories, both views): comparator
Recall@5 (six questions, $d=8$), paired difference and one-sided 97.5\% bound (margin $-0.05$).}
\label{tab:ni}
\input{generated/ni_compact}
\end{table}
\begin{figure}[t]\centering
\includegraphics[width=257.33461pt]{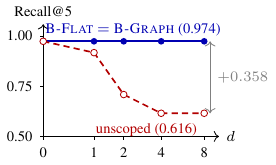}
\caption{Recall@5 of the temporal-update question at distractor dose $d$
(descriptive; 95\% intervals within $\pm\DoseHalfWidth$).}
\label{fig:dose}
\end{figure}
\else
\begin{figure}[t]
\begin{minipage}[t]{0.50\linewidth}\centering\small\vspace{0pt}\setlength{\abovecaptionskip}{0pt}\setlength{\belowcaptionskip}{6pt}
\captionof{table}{Entitled-recall non-inferiority (10{,}000 histories, both views): comparator
Recall@5 (six questions, $d=8$), paired difference and one-sided 97.5\% bound (margin $-0.05$).}
\label{tab:ni}
\input{generated/ni_compact}
\end{minipage}\hfill
\begin{minipage}[t]{0.47\linewidth}\centering\vspace{0pt}
\input{generated/dose_figure}
\captionof{figure}{Recall@5 of the temporal-update question at distractor dose $d$
(descriptive; 95\% intervals within $\pm\DoseHalfWidth$).}
\label{fig:dose}
\end{minipage}
\end{figure}
\fi

\subsection{Wrong-principal competition (H3) and the joint decision}
\label{sec:results-wrong-principal}

\input{sections/06_h3_final}

\subsection{Derivation and native validation (H4)}
\label{sec:results-derivation}
\label{sec:results-native}

Transformation and mutation tests exercise H4a, H4c and H4d case by case;
in the matched H4b study, derived items stayed available under intersection in
8 of 8 cases and in none under blanket suppression
(App.~\ref{app:supporting-full}). In the native frame, AI review marked 195 of 210 answers correct in 42 scenarios (four unfavorable cases in App.~\ref{app:native-full}), and a scoped native qualification passed all 39 phases.

\paragraph{Outcome accounting.}\label{sec:results-accounting}\label{sec:results-supporting}
\label{sec:results-negative}

All 10{,}000 histories, 480{,}000 retrieval cells and 30{,}000 planned answers are
retained. Under the original protocol, 85 answers
(35/34/16 for \bflat{}/\bgraph{}/unscoped) failed the citation contract; the amended view recovered 67, leaving 18 in ten histories (App.~\ref{app:accounting}). {} {} 
\unskip

%% file: generated/ni_compact.tex
\setlength{\tabcolsep}{4pt}\begin{tabular}{@{}lccc@{}}
\toprule
Comparator & R@5 & Mean $\Delta$ & 97.5\% bound \\
\midrule
Unscoped & 0.674 & $+0.298$ & $+0.2900$ \\
Session silo & 0.740 & $+0.233$ & $+0.2283$ \\
Bounded post-filter & 0.972 & $0.000$ & $-0.0020$ \\
Principal postings & 0.972 & $0.000$ & $-0.0020$ \\
\midrule
\bflat{}, \bgraph{} & 0.972 & \multicolumn{2}{c}{all 8 pass} \\
\bottomrule
\end{tabular}

%% file: generated/dose_figure.tex
\includegraphics[width=132.555pt,height=79.13074pt]{figures/dose-recall.pdf}

%% file: sections/06_h3_final.tex
The Holm tests confirm recall superiority over unscoped retrieval (\HRecallEstOrig) and an
advantage that grows with distractor dose (H3c slope \HSlopeFlatOrig{} on $\log_2(d{+}1)$) in both views; both are
retrieval-only (Table~\ref{tab:h3}; Figure~\ref{fig:holm} in App.~\ref{app:accounting}).
Figure~\ref{fig:dose} shows the mechanism: unscoped Recall@5 falls from \DoseUnscZero{} to
\DoseUnscEight{} as distractors about other participants displace entitled evidence, whereas both
architectures hold \DoseArch{} because no distractor reaches their candidates: in all 10{,}000
histories they fetched the same 20 candidates and context at every dose, and unscoped retrieval did so in none.
Neither architecture produced a wrong-principal substitution: \bflat{}
\HSubFOrig{} of \HJudFOrig{} and \bgraph{} \HSubGOrig{} of \HJudGOrig{} judged answers, against
\HSubUOrig{} of \HJudUOrig{} for unscoped retrieval (amended view: \HSubFAmend{}, \HSubGAmend{} and \HSubUAmend{} of
\HJudGAmend{}--\HJudUAmend{}).  At this
base rate the frozen test cannot confirm the reduction (App.~\ref{app:accounting}), so H3a/b is an underpowered null, not evidence
against the mechanism; an exploratory sign test (not prespecified) finds all \HSignFor{} discordant pairs
favoring each architecture (one-sided $p=\HSignP$); the joint decision is \emph{\HJointSummary}.

\begin{table}[t]
\centering
\small
\caption{Six-member Holm family ($\alpha=0.05$), worst-case missingness supplement.
Substitution: identification range of the unscoped-minus-target difference, upper $p$ ($^\dagger$).}
\label{tab:h3}
\input{generated/h3_family}
\end{table}

%% file: generated/h3_family.tex
\setlength{\tabcolsep}{4pt}\begin{tabular}{@{}lcccccc@{}}
\toprule
& \multicolumn{3}{c}{Original view} & \multicolumn{3}{c}{Amended view} \\
\cmidrule(lr){2-4}\cmidrule(lr){5-7}
Holm member & Estimate & $p$ & Rej. & Estimate & $p$ & Rej. \\
\midrule
Recall, \bflat{} & $+0.298$ & $<10^{-6}$ & yes & $+0.298$ & $<10^{-6}$ & yes \\
Recall, \bgraph{} & $+0.298$ & $<10^{-6}$ & yes & $+0.298$ & $<10^{-6}$ & yes \\
Substitution, \bflat{} & [$-0.0028$, $+0.0023$] & $1$$^\dagger$ & no & [$+0.0000$, $+0.0010$] & $1$$^\dagger$ & no \\
Substitution, \bgraph{} & [$-0.0027$, $+0.0023$] & $1$$^\dagger$ & no & [$-0.0001$, $+0.0010$] & $1$$^\dagger$ & no \\
Dose slope, \bflat{} & $+0.130$ & $<10^{-6}$ & yes & $+0.130$ & $<10^{-6}$ & yes \\
Dose slope, \bgraph{} & $+0.130$ & $<10^{-6}$ & yes & $+0.130$ & $<10^{-6}$ & yes \\
\midrule
Joint decision & \multicolumn{3}{c}{not established} & \multicolumn{3}{c}{not established} \\
\bottomrule
\end{tabular}

%% file: sections/07_limitations.tex
\section{Limitations}
\label{sec:limitations}

The guarantee is only as strong as its trusted base: a misresolved transport identity yields a
wrong decision that the oracle executes faithfully, so identity resolution is where a deployment
must invest; the identity study (App.~\ref{app:supporting-full}) shows that every fail-closed policy
removes exposure at a measured cost in entitled recall. The property is read-audience
confidentiality at the context boundary; full selection noninterference additionally requires
per-audience index statistics (\S\ref{sec:guarantee}). Confirmatory inference covers one controlled
generator, one answer model and one lexical ranker, and the external corpora establish recall parity
without cross-audience ground truth. Semantic outcomes are AI judgments without the approved
consistency audit, and external answers mix three generation models, so neither is independent or
human validation. Native evidence is descriptive, and its historical conformance receipts predate the
evaluated revision.

%% file: sections/08_conclusion.tex
\section{Conclusion}
\label{sec:conclusion}
\unskip

Authorization before context extends across the memory lifecycle: no forbidden item entered
any exact context, recall matched policy-equivalent baselines at every $k$ and
increasingly exceeded unscoped retrieval under distractors, and no architecture substituted a wrong
principal, although unscoped substitutions were too rare to establish the joint decision.
Confidentiality in agent memory is therefore a property of the memory architecture rather than of
model behavior: binding audiences at write time and admitting before ranking costs no recall and
removes the slot competition that degrades unscoped memory, with a guarantee that holds on the exact
context of every model attempt. 
\paragraph{Future work.} The same boundary discipline extends from confidentiality to integrity
and action. Origin-bound authority would admit content not authored by the owner or an explicitly
delegated proxy only as attributed, quarantined data, never as instructions, trusted memory or
effect authority, containing prompt-injection damage by construction rather than by detection \citep{tmanm2026,debenedetti2025camel,costa2025fides}. An
effect gateway would commit external actions only through deterministic, source-closed, single-use
authorizations bound to exact arguments, so a hijacked model can at most propose an effect. Task
runtimes for open-ended research, browsing and code execution would be treated as compromised, their
outputs reaching memory or effects only through these mediated boundaries. Within the present design,
per-audience index statistics would add full selection noninterference, and substitution studies at
higher base rates would resolve H3a/b.

%% file: sections/09_statements.tex
\subsection*{AI use statement}

Generative AI was used substantially in this research and manuscript preparation. AI coding assistants were used to implement the reference architectures, the native-runtime changes, the evaluation harness, the analysis code and tests, under the author's specifications and review; they also drafted
and reviewed specifications, experimental and statistical designs, and performed
technical reviews of implementation and analysis packets. AI judgments are part
of the evidence: pre-confirmation answer review combined approved
AI-assisted judgments with ten judgments by a study author; the native answers were labeled by
the executing AI session; and confirmation answers were judged by AI judging sessions on several recorded
routes{} (the approved AI consistency audit was not run before
submission){} (Section~\ref{sec:stats} and
App.~\ref{app:judging}), whose limits we report. AI coding sessions also operated, checked and admitted the final judging windows
under study approval. AI assistants also helped formulate the formal model and its mathematical claims, drafted
the proofs, interpreted results, created figures, performed literature search, citation
verification and reference formatting, and drafted and revised this manuscript. The synthetic
histories were generated by AI-assisted generator code whose language is rendered by
deterministic templates without model calls. The evaluated answer models, including the three external generation models,
are subjects of study, not author aids. OpenAI Codex also assisted with preprint formatting and source-package checks. The author
directed the research, claims and limitations and takes responsibility for the final content, including text, claims and artifacts produced with generative AI.

\subsection*{Ethics statement}

All confirmatory and native experiments use synthetic histories and run in fresh
isolated research state; no private
conversation, real participant identity or production trace is used or released.
Independently authored corpora are used under their recorded licenses with disclosed adaptations: LongMemEval (MIT), EverMemBench (Apache-2.0) and GateMem (CC-BY-4.0); LoCoMo (CC-BY-NC-4.0) and GroupMemBench (no asserted license) are not redistributed \unskip. The mechanism restricts what memory reaches a model
and does not create new capabilities for misuse; a misconfigured identity
resolver could, however, grant memory to the wrong audience, which is why the
guarantee is stated conditionally and unresolved viewers fail closed.

\subsection*{Reproducibility statement}

Section~\ref{sec:eval} and Appendix~\ref{app:protocol} specify the frozen
protocol, sample plan, model and embedding pins, decision rules and
answer-state taxonomy. The research artifact supplied for double-blind review accompanies this preprint as ancillary
material (CC BY-NC 4.0 for project-authored material; third-party notices preserved).
It contains the policy oracle, the independent \bflat{} and \bgraph{} stores and the
evaluation-only baselines, the controlled generator, the conformance suites, 365 tests and
the analysis code, with a claim-to-evidence map naming each table's and figure's command,
inputs and expected output. One command, \texttt{reproduce\_all.sh}, runs every check from a
clean copy without model calls: the generator regenerates the controlled scenarios
byte-identically; the conformance replay reproduces 94 of 94 differential cases, 71 killed
mutants and 24 transformation cases; the supporting replay reproduces all four tables; and a
headline replay regenerates every headline number from per-history reductions with the
frozen estimator code (H1 counts, Tables~\ref{tab:ni} and~\ref{tab:h3} in both views,
Figures~\ref{fig:dose} and~\ref{fig:topk}, substitution counts, the sign test and slot
noninterference), checking each value exactly against the submission run. The native source
slice and the large per-history outputs and judgment records are not included in this
preprint artifact; their release is deferred to the final version. Model-backed commands for fresh stochastic replication are documented separately
and require the reader's own model access.

%% file: sections/A_formal.tex
\section{Proofs}
\label{app:formal}

Notation follows \S\ref{sec:model}. Fix an authoritative snapshot at time $t$.

\paragraph{P1.} Immediate from Eq.~(\ref{eq:policy}): membership in
$\mathrm{Adm}(C,t)$ requires some $A\in\AuthSets(m,t)$ with $V(C)\subseteq A$.
Hence no item reaches a viewer outside every authorized audience. \hfill$\square$

\paragraph{P2.} If $V(C_1)\subseteq V(C_2)$ and $m\in\mathrm{Adm}(C_2,t)$, the
witness $A$ with $V(C_2)\subseteq A$ also satisfies $V(C_1)\subseteq A$, and
currentness does not depend on $C$; so $m\in\mathrm{Adm}(C_1,t)$. Adding a
current grant to $\AuthSets(m,t)$ can only add admitting containers. \hfill$\square$

\paragraph{P3.} With unresolved viewers, $V(C)=U\ni\star$. Owner-private and shared
audiences exclude $\star$ (\S\ref{sec:labels}), and so does every grant, which
names resolved participants only; so $V(C)\subseteq A$ holds only for
$A=\mathrm{Aud}(m)=U$, and
$\mathrm{Adm}(C,t)=\{m:\mathrm{Aud}(m)=U\wedge\mathrm{Current}(m,t)\}$.
Implementations store the three audience kinds as typed labels and an unresolved
viewer state as $V(C)=U$, so this is exactly the rule ``unresolved viewers see
public items only''.
\hfill$\square$

\paragraph{P4.} Write-time binding sets $\mathrm{Aud}(m')=A$ for an injected
item $m'$, where $A$ is the set of resolved viewers of its origin container, independent of
content; a write whose origin viewers cannot be determined is refused. A grant requires an authenticated, object-specific
decision that content cannot mint, and derivation reads $\mathrm{Aud}$ only.
By P1, $m'$ reaches only containers with $V(C)\subseteq A$ unless such a decision
is made. This is a confidentiality property, not an integrity guarantee. \hfill$\square$

\paragraph{P5.} For an intersection-derived item $d$ with sources $S$,
$\mathrm{Aud}(d)=\bigcap_{s\in S}\mathrm{Aud}(s)\subseteq\mathrm{Aud}(s)$ for
every $s$; $o\in\mathrm{Aud}(d)$ because $o$ belongs to every audience, and
$\star\in\mathrm{Aud}(d)$ only if every source is public. Any
$V(C)\subseteq\mathrm{Aud}(d)$ therefore satisfies $V(C)\subseteq\mathrm{Aud}(s)$
for every source. Because derivation reads $\mathrm{Aud}$ and never
$\mathrm{Grants}$, no source grant enters $\mathrm{Aud}(d)$. \hfill$\square$

\paragraph{Proposition~\ref{prop:guarantee}.} The canonical basis selector
$\sigma(m,C,t)$ enumerates the members of $\AuthSets(m,t)$ separately and
returns $\mathrm{Aud}(m)$ if $V(C)\subseteq\mathrm{Aud}(m)$, otherwise the
canonical minimum grant audience containing $V(C)$, and $\bot$ if none exists.
The implementation admits $m$ iff $\mathrm{Current}(m,t)$ and
$\sigma(m,C,t)\neq\bot$. A caller-supplied binding that is invalid, stale or
swapped is rejected even when another valid binding exists; that is a fail-closed error, not evidence of inadmissibility, and is tested
separately; part (ii) below concerns requests without such an error.

\emph{(i)} By complete mediation, $m$ entered the attempt only through this
check against the snapshot at $t$, so $\mathrm{Current}(m,t)$ holds and
$\sigma(m,C,t)=A\neq\bot$ with $A\in\AuthSets(m,t)$ and $V(C)\subseteq A$; hence
$m\in\mathrm{Adm}(C,t)$.
\emph{(ii)} If $m\in\mathrm{Adm}(C,t)$, some $A\in\AuthSets(m,t)$ contains
$V(C)$; since $\sigma$ enumerates every member of $\AuthSets(m,t)$ it returns an
authorizing audience, so the implementation admits $m$, and an exhaustive backend
returns every admitted item of $Q(C)$. \hfill$\square$

%% file: sections/B_protocol.tex
\section{Frozen Protocol, Mutations and Deviations}
\label{app:protocol}

\subsection{Frozen protocol summary}

\begin{table}[!ht]
\centering
\small
\caption{Frozen confirmation protocol.}
\label{tab:protocol}
\begin{tabular}{P{0.34\linewidth}P{0.58\linewidth}}
\toprule
Item & Frozen value \\
\midrule
Confirmation histories & 10{,}000 (draws 40--10{,}039), eight themes $\times$ 1{,}250 \\
Retrieval cells & 48 per history; 480{,}000 total; one deterministic replicate \\
Answers & 1 per history for unscoped, \bflat{}, \bgraph{}; 30{,}000 planned \\
Attempts & $\le 2$ identical-input attempts per answer; $\le 330$ per 100-history chunk; $\le 33{,}000$ total \\
Retrieval & shared deterministic lexical index per history (FTS5 BM25, reciprocal-rank fusion); fetch budget 20; $k=5$; ties by item identifier \\
Answer model & \texttt{gpt-5.6-luna}, medium reasoning, no fallback, observed-identity check \\
Embedding & \texttt{text-embedding-3-small} pinned (1{,}536 dims, unit-$\ell_2$, cosine); not used by confirmation ranking \\
Primary endpoint & paired authorized Recall@5 difference \\
Non-inferiority & margin $-0.05$; one-sided empirical-Bernstein bound, $\alpha=0.025$; all 8 must pass \\
Superiority & 6 contrasts after the gate; Holm familywise $\alpha=0.05$ \\
H3c & paired OLS slope of $\delta(d)$ on $\log_2(d+1)$, $d\in\{0,1,2,4,8\}$ \\
Stopping & no interim outcome looks; no optional stopping \\
Missing outcomes & no imputation, substitution or silent exclusion \\
Pilot & 32 held-out histories, 5 answer replicates, excluded from confirmation \\
Planned joint power & min.\ simultaneous MC lower bound 0.935 over 40 scenarios (target 0.90) \\
Arm order & seeded Latin rotation \\
Execution & subscription transport; 3 shared workers; no paid fallback \\
Extraction fitness & 20 episodes vs.\ human annotation: entity recall 1.0, relation recall 28/29, precision 1.0 \\
\bottomrule
\end{tabular}
\end{table}

 \unskip

\subsection{Mutation operators}
\label{app:mutations}

Common operators: reverse the subset relation; treat unknown viewers as
owner-private; omit one current viewer; add one unauthorized audience; bypass
authorization on direct lookup; merge derived audiences by union; admit
incomplete-provenance summaries; let a source grant authorize a new persisted
derivation or let a derived artifact inherit it; skip endpoint cascade on graph
edges; remove an owner-policy branch; let query order select the reason code;
reduce bounded over-fetch compensation; union two separately approved grants;
reject a valid grant on an owner-private item or mutate the origin label when
applying it;
let a model generate or rewrite an authoritative identifier or source closure;
authorize after ranking; change the sealed context after its snapshot.
Native-only operators additionally target write-producer inventory, identity-only
entity reuse, stale currentness or authorization decisions, snapshot placement
relative to capacity waiting and retries, post-transport currentness reads,
binding forgery, duplicate authorization-registry entries, realm and audience partitioning,
projection collapse and bounded refill.
\unskip

\subsection{Amendments and deviations}
\label{app:deviations}

\begin{enumerate}
\item \textbf{Pre-confirmation human review.} The originally required
concealed-repeat or independent-adjudicator reliability check was replaced, by a
prospective amendment recorded before confirmation, with study-approved
AI-assisted judgments plus ten direct judgments by a study author without arm or system
labels (descriptive agreement 10/10; repeat concealment not established). The
amendment changed no hypothesis, population, sample, margin, multiplicity rule,
power target or success rule.
\item \textbf{Source adjudication.} One LoCoMo question whose target is a named
vehicle and whose retained answer abstained is recorded as an answer miss, not a
wrong-principal substitution; its retrieval recall is unscorable because the
original question has no evidence annotations. This was a post-review source
adjudication, not a blinded rating.
\item \textbf{Extraction fitness.} An earlier fitness run failed; a bounded
20-episode replacement passed. The failed run and its charges are retained and
receive no fitness credit.
\item \textbf{Pilot.} The pilot was not rerun after later code repairs; it is
reported historically at its original revisions and excluded from confirmation.
\item \textbf{Citation recovery.} After the original protocol completed, a
separately recorded amendment allowed a recovery attempt for original terminal
citation-contract failures. Both the original-protocol and the amended views are
reported; neither is selected as primary. \unskip
\item \textbf{Analysis correction.} A comparison in the analysis adapter
required completion publication order to equal physical attempt order and
thereby rejected three valid answers of one history. The corrected analysis
compares completed records by unique observation identity while preserving
full-content equality; evidence-bound revalidation then resolved one of the three
answers and left two awaiting review (both since judged). The other 9{,}999 histories and all 10{,}000
retrieval reductions are byte-identical before and after, no answer was
regenerated, no admission mismatch remains, and an independent technical review
accepted the correction.
\item \textbf{Judging route.} For semantic answer review, an approved study amendment
restricted to this phase replaced the requirements that judging use the pinned model and that the answer model never be the sole judge
with AI judging and no human assessors; unavailable model metadata is recorded as
unknown. It does not amend any other call class. Judging began in one AI
coding-assistant session; a later approved override moved it to parallel judge
sessions (\texttt{gpt-6-sol}, low reasoning effort) on disjoint batches under one
coordinating session (\texttt{gpt-6-astra}, extra-high effort). A judge context
that reaches its recorded context limit receives no further work; each new \texttt{gpt-6-sol} context is qualified individually, and no judge context has
seen keys or prior labels. A later study-approved switch moved the remaining queue to fresh-session
\texttt{claude-opus-5-5} judges at low effort, which share one procedure-level
qualification (18 controls) reused unchanged across windows; each route is recorded as
a provenance stratum. When the validator rejected one internally inconsistent
judgment, the queue stopped by design; an study-approved successor with the same
judges, rubric and validators resumed it from the last durable snapshot, with its production checks done by the executing AI coding session under study
approval (neither independent nor human), and the 17
subjects of the rejected batch were judged once more (the rejected response is
retained). Parked subjects were then closed in one final pass: 170 were admitted from existing saved judgments (86 previously validated; 84
validated by the unchanged validator at admission) and 107 received one fresh
judgment, leaving none pending; both successor windows were checked the same way. Rubric,
schema and citation validation are unchanged, and earlier saved judgments are
retained.
\item \textbf{Audit scope.} The study leads approved an AI consistency audit by the same
judging route (App.~\ref{app:judging}) in place of independent semantic validation{}
(not run before submission; \S\ref{sec:stats}){}, and
required that this limitation be reported.
\item \textbf{Recovery hold.} After 67 of the 85 original failures were
recovered, further recovery calls were put on hold; the 18 remaining identities
are retained.
\item \textbf{Missingness supplement.} After the recovery hold, a lightweight
worst-case supplement over compatible completions of genuinely missing binary
outcomes was conditionally authorized; it keeps the shared unscoped outcome, the
eight gates and the full Holm family, is reported separately from the frozen unavailable result, and  was
executed once all labels were saved (\S\ref{sec:results-wrong-principal}).
\item \textbf{External answer models.} External answers were generated under three
recorded model strata (App.~\ref{app:external}) instead of the single pinned
model, and eight valid responses recorded outside the run journal leave a
selection question open: for seven identities the earliest valid completion
differs from the journal's recorded choice. Both selections are retained and
neither is promoted; external results are reported per stratum.
\end{enumerate}

%% file: sections/C_accounting.tex
\section{Outcome Accounting}
\label{app:accounting}

\paragraph{Retrieval.} Planned retrieval cells per arm: 100{,}000 each for
unscoped, \bflat{} and \bgraph{}; 60{,}000 each for silo, bounded post-filter and
principal postings; 480{,}000 in total. All 10{,}000 per-history retrieval
reductions are complete and identical across the two answer views.

\paragraph{Answers.} Table~\ref{tab:answer-states} gives per-arm states.

\begin{table}[!ht]
\centering
\small
\caption{Answer states per arm (10{,}000 planned answers each; generated from the accepted export, excluding the review overlay). \emph{Pending}
means contract-valid but not yet semantically reviewed; it is not a correctness
judgment; every pending answer was judged afterwards
(\S\ref{sec:results-wrong-principal}).}
\label{tab:answer-states}
\input{generated/answer_states}
\end{table}
 No answer is in the \emph{disputed}, \emph{exact-value
contradiction} or \emph{admission failed} state in either view. Under the
original protocol, 60 histories contain at least one terminal failure; under the
amended view, 10 histories contain at least one unresolved recovery. A history can
contain answers in several states.

\paragraph{Physical attempts.} 30{,}524 historical physical answer attempts and
their charges are retained unchanged. \unskip

\paragraph{Missing answers.} The original view has 85 missing answers in 60
histories (35/34/16 by \bflat{}/\bgraph{}/unscoped); the amended view has 18 in
10 histories (7/8/3). All 18 lie in the attribution-conflict slot at dose 8 and
in the single frozen six-question stratum; by source theme they are care visit 9,
client design 3, family travel 3, volunteer shift 2 and housing inspection 1.
These counts support no missing-at-random assumption. Pending semantic reviews
are existing responses, not missing answers. Further answer recovery is on hold.

\paragraph{Missingness supplement.} The worst-case supplement (conditionally authorized after the
recovery hold; App.~\ref{app:deviations}, item 10) varies, for each view, only the genuinely missing binary substitution
outcomes over every compatible completion, keeps the one shared unscoped outcome per
history, and reapplies the frozen bound, eight gates and six-member Holm family
through integer sufficient statistics (at most eight completions per history; a
bounded search for the original 85). Pending reviews, disputes and contradictions are
not treated as missing. It is reported alongside, never instead of, the frozen
unavailable result \unskip.

\paragraph{Why H3a/b cannot be rejected.} With $n$ complete histories the
empirical-Bernstein bound subtracts at least the range term
$7r\ln(2/\alpha)/(3(n-1))$; with $r=2$, $\alpha=0.05$ and $n=9{,}954$ this is about
$0.0017$, larger than the complete-case mean substitution difference of about
$7/9{,}954\approx0.0007$. The test therefore cannot reject even before the missing
outcomes are bounded, and the worst-case envelope $p$ is 1 in both views.{}

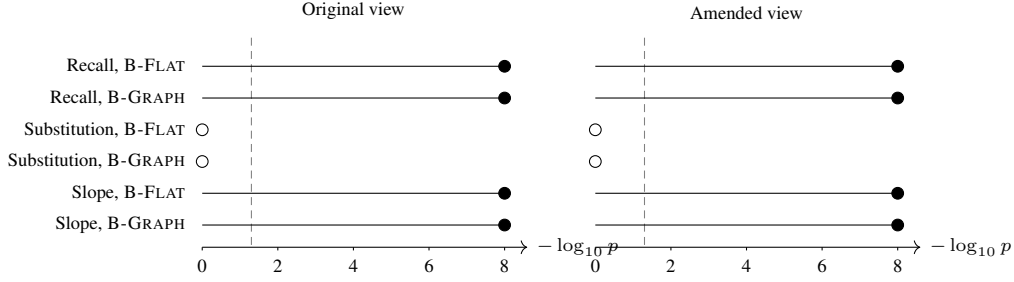
\begin{figure}[h]
\centering
\input{generated/h3_forest}
\caption{Holm family per view: $-\log_{10}$ of each member's $p$-value (upper
envelope for substitution); filled markers are Holm rejections, and the dashed
line marks $p=0.05$.}
\label{fig:holm}
\end{figure}

%% file: generated/answer_states.tex
\begin{tabular}{lrrrrrr}
\toprule
& \multicolumn{3}{c}{Original protocol} & \multicolumn{3}{c}{Amended recovery} \\
\cmidrule(lr){2-4}\cmidrule(lr){5-7}
Arm & Resolved & Pending & Terminal & Resolved & Pending & Unresolved \\
\midrule
\bflat{} & 4{,}812 & 5{,}153 & 35 & 4{,}825 & 5{,}168 & 7 \\
\bgraph{} & 4{,}807 & 5{,}159 & 34 & 4{,}815 & 5{,}177 & 8 \\
Unscoped & 5{,}103 & 4{,}881 & 16 & 5{,}113 & 4{,}884 & 3 \\
\midrule
Total & 14{,}722 & 15{,}193 & 85 & 14{,}753 & 15{,}229 & 18 \\
\bottomrule
\end{tabular}

%% file: generated/h3_forest.tex
\includegraphics[width=384.3968pt,height=108.45982pt]{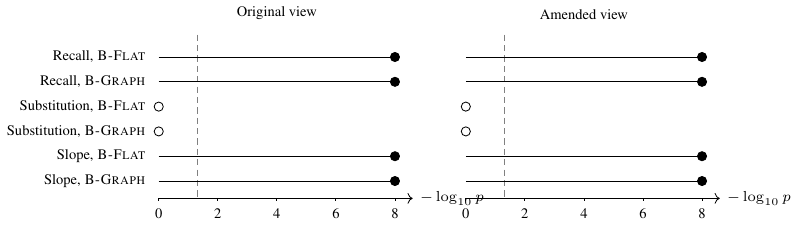}

%% file: sections/D_related_matrix.tex
\section{Extended Related-Work Matrix}
\label{app:related}

Table~\ref{tab:related} summarizes what each line of work establishes and how this paper differs.

\paragraph{Authorization before retrieval.} Authorization-First Retrieval
\citep{afr2026} argues, for multi-agent retrieval-augmented generation over
documents, that authorization must precede any learned component and reports
that retrieve-then-filter pipelines expose unauthorized content structurally.
Filtered vector search is a known systems trade-off in which attribute filters
break the recall assumptions of approximate indexes \citep{chronis2025filtered}.
We adopt both lessons and do not claim them; our setting adds persistent writes
whose audiences are derived from the conversation transport, transformations
whose labels must be computed, overlapping audiences, declassification, and a
recall study against a policy-equivalent principal-postings baseline.

\paragraph{Governed and multi-user agent memory.} Collaborative Memory
\citep{rezazadeh2025collaborativememory} shares memory among users and agents
through time-varying access graphs, provenance-tagged fragments and read/write
policies that project filtered views. \citet{yang2026multiuser} formalize
multi-user agents as a multi-principal decision problem and observe privacy
violations that grow over multi-turn interaction. Governed shared-memory systems
include a multi-tenant fleet-memory service whose authors report, among other
failure modes, a sub-tenant scope bypass on a get-by-identifier path
\citep{memclaw2026}, and a local-first memory system with scope isolation,
role-based access and verified erasure \citep{superlocalmemory2026};
\citet{chen2026deploymentmemorization} treat deployment-time memorization as a
privacy--utility frontier and observe residue of deleted raw data in derived
tiers. Relative to these, our audiences are derived from the transport identities of the participants present
rather than configured, derivation and direct lookup are mediated by the same
closed rules, and the endpoint is the exact assembled context of each physical
model attempt.

\paragraph{Memory privacy benchmarks.} CIMemories
\citep{mireshghallah2025cimemories} measures whether models control
information flow from persistent memory by task context and finds violations
that accumulate across tasks; PiSAs \citep{pisas2026} measures cross-user
spillage through outputs, inter-agent messages and memory; GateMem
\citep{gatemem2026} scores utility, access control and forgetting jointly for
multi-principal shared memory; GroupMemBench \citep{yang2026groupmembench} and
EverMemBench \citep{evermembench2026} evaluate multi-party and long-horizon
memory utility. Most of these score model output or task utility; we score the assembled context directly and use independently authored corpora under a disclosed audience overlay,
reporting GateMem's missing audience labels as a construct limitation
(\S\ref{sec:eval-data}; Table~\ref{tab:coverage}).

\paragraph{Authority, integrity and provenance laundering.} A separate line of
work concerns what memory authorizes an agent to \emph{do}, or whether memory
can be trusted. Consolidation can erase the source constraints of a retained
claim \citep{authmembench2026}; memory writers can invent authority that
executors act on \citep{eal2026}; consolidation can rewrite external
observations as user history \citep{ppmf2026}; and poisoning defenses based on
content or lineage can be unsound under laundering, motivating write-time,
origin-bound, non-malleable authority \citep{tmanm2026}. MAP-Graph
\citep{mapgraph2026} separates hard permission eligibility from graded path
trust and risk-gated actions; MemSecBench \citep{memsecbench2026} measures the
persistence and consequences of memory poisoning; MutMem \citep{mutmem2026}
makes memory mutations cryptographically authorized. We share the write-time
origin-binding principle but address a different property (read-audience confidentiality from source to
context) and claim no action-authority,
integrity or poisoning-defense result.

\paragraph{Information flow and declassification.} Our labels descend from
lattice and decentralized information-flow control
\citep{bell1976secure,denning1976lattice,myers1997decentralized}, our grants
from the declassification literature \citep{sabelfeld2009declassification},
and our derived-data rules from provenance-based access control and derived-data
policy \citep{bates2013provenance,denhartog2016datafusion}. Agent-level
information-flow systems such as FIDES \citep{costa2025fides} and CaMeL
\citep{debenedetti2025camel}, and the reference-monitor framing of FORGE \citep{palumbo2026forge},
target control and effect flows under prompt injection. Contextual integrity
\citep{nissenbaum2009privacy} motivates the setting; we implement
only its recipient/audience slice, not information-type or
transmission-principle norms.

\begin{table}[h]
\centering
\scriptsize
\caption{Related work: what each line establishes and how this paper differs.}
\label{tab:related}
\renewcommand{\arraystretch}{1.2}
\begin{tabular}{@{}P{0.26\linewidth}P{0.33\linewidth}P{0.33\linewidth}@{}}
\toprule
Work & What it establishes & Distinction \\
\midrule
AFR \citep{afr2026} & Authorization before learned retrieval for multi-agent RAG & Persistent writes, derived labels, overlapping audiences, grants, recall study \\
Collaborative Memory \citep{rezazadeh2025collaborativememory} & Access graphs, provenance-tagged fragments, filtered views & Participation-derived audiences; closed derivation; exact-context endpoint \\
GateMem \citep{gatemem2026} & Joint utility, access-control and forgetting benchmark & No checkpoint has authoritative audience labels; reported as a construct limitation \\
CIMemories \citep{mireshghallah2025cimemories} & Contextual-integrity violations accumulate across tasks & Deterministic exclusion before the model; recall measured \\
PiSAs \citep{pisas2026} & Cross-user spillage across surfaces & Structural memory-to-context exclusion \\
AuthMem-Bench \citep{authmembench2026} & Authority collapse at consolidation & Read confidentiality, not authority to act \\
TMA-NM \citep{tmanm2026} & Origin-bound non-malleable authority against poisoning & Shares write-time origin binding; different property \\
EAL-Bench \citep{eal2026}, PPMF \citep{ppmf2026} & Memory-created or laundered action authority & No action-policy claim \\
MAP-Graph \citep{mapgraph2026} & Eligibility + path trust + action gating & Exact audiences, lifecycle, recall study \\
MemSecBench \citep{memsecbench2026}, MutMem \citep{mutmem2026} & Poisoning persistence; authorized mutation & Integrity is out of scope \\
Governed memory \citep{memclaw2026,superlocalmemory2026} & Multi-scope governed memory services & Transport-derived audiences; derivation; mediation \\
FIDES, CaMeL, FORGE \citep{costa2025fides,debenedetti2025camel,palumbo2026forge} & Agent information flow and reference monitors for actions & Narrow read-audience property at the memory boundary \\
IFC lineage \citep{bell1976secure,denning1976lattice,myers1997decentralized,sabelfeld2009declassification} & Labels, lattices, declassification & Agent-memory lifecycle specialization only \\
Provenance policy \citep{bates2013provenance,denhartog2016datafusion} & Provenance-based and derived-data access control & Specialized to relationship audiences and exact context \\
Zanzibar \citep{pang2019zanzibar} & Relationship-based authorization & Userset containment only \\
Filtered ANN \citep{chronis2025filtered} & Filters break index recall assumptions & Measured, not claimed \\
\bottomrule
\end{tabular}
\end{table}

%% file: sections/E_details.tex
\section{Implementation and Evidence Details}
\label{app:details}

This appendix adds implementation and evidence detail only; definitions, the
threat model and non-goals are in \S\ref{sec:model}.

\subsection{Stored fields and provenance}

When one artifact is stored as several physical copies (an origin copy and grant copies), the
copies count as one item in every arm. Every recorded write persists canonical participant identifiers, the origin
audience $\mathrm{Aud}(m)$ as a typed label, the transport and container
identifiers, the identity-resolution confidence and evidence class, source
identifiers, the write timestamp and policy version, and a stable fact-bearing
object identity to which later grants link. Provenance is \emph{code-bound}: a
trusted route adapter mints each authoritative source identifier from an immutable
input record, and derived source sets are constructed from the records a
transformation actually consumed. Models may generate or transform content, but
never generate, select, parse or rewrite an authoritative source, audience, grant
or derivation identifier; output with missing or additional provenance is
suppressed or quarantined. Shared audiences use one canonical digest over a typed,
sorted member projection; a mismatch in kind, digest, member count or members is a
security failure, not a lookup miss.

\subsection{Revocation and narrowing}
\label{app:revocation}

Three changes are simple under
Eq.~(\ref{eq:policy}), though we do not evaluate them. Revoking a grant deletes one
grant record, so later attempts no longer see it in $\mathrm{Grants}(m,t)$.
Deleting or revoking an item makes it, and every item derived from it,
non-current. Removing a member deletes that participant from every stored shared audience and
grant that contains it; this can only shrink the set of admitting
containers, and no item needs to be deleted or merged, because every audience
still contains $o$. Each change takes effect from the next attempt's snapshot.

\subsection{Native realization mechanisms}
\label{app:native-mech}
Three mechanisms required native-specific engineering:
(i)~\emph{projection collapse}: origin and grant projections of one artifact
share an \emph{authorization subject identity}, consume at most one final result
slot, and are refilled from the already-authorized candidate set with a bounded,
deterministic schedule that returns fewer than $k$ results rather than broaden
the query;
(ii)~\emph{identity-only entity reuse}: graph entity nodes may be reused across
audiences for identity resolution but never carry factual authority, and only an
explicit grant over an existing fact-bearing artifact creates a destination
projection; and
(iii)~\emph{attempt admission}: final source and authorization reads for each
physical attempt use one private read-only snapshot, and response attribution
after transport uses only immutable continuation checks, never a second
currentness read. 
\subsection{Additional evidence sources}
\label{app:external}
\paragraph{Independently authored external evidence.} Retrieval preparation covers
all 2{,}549 questions and 15{,}294 cells, and all 7{,}647 planned answer identities
(1{,}500 LongMemEval, 5{,}943 LoCoMo, 204 EverMemBench across the three answer
arms) have a recorded contract-valid or explicitly reused output. That is completion of generation, not a finding of accuracy: semantic grading has not been done.
The stored retrieval scores of all 15{,}294 external cells show every arm retrieving
identically: no arm, including unscoped retrieval, placed a forbidden item in context,
and entitled Recall@5 is equal across all six arms (0.917 on 470 recall-eligible
LongMemEval, 0.468 on 1{,}531 LoCoMo and 0.090 on 68 EverMemBench questions). The overlay
leaves these corpora without competing cross-audience evidence, so they show recall
parity on independently authored data rather than exclusion. The outputs come from three recorded generation models
(Table~\ref{tab:external}); within each corpus the arms have similar model mixes,
but 140 questions have different models across their arms, so their paired
comparisons are model-confounded, and the corpora differ sharply in model mix.
Accounting holds 8{,}015 records: 5{,}190 subscription physical attempts and
2{,}825 command-line reservations whose underlying send count is unknown; they
are not all verified physical requests. Twelve historical answers were adopted
without their earlier reviews. For eight valid responses recorded outside the run
journal, the earliest valid completion differs from the recorded choice for seven
identities, and the full response differs for five; both selections are retained
and the scientific choice between them is unresolved
\unskip.

\begin{table}[h]
\centering
\small
\caption{External answer identities by corpus and recorded generation model
(three arms per question; generated from the pinned import). Mixed-model q.:
questions whose three arms were answered by more than one model.}
\label{tab:external}
\input{generated/external_strata}
\end{table}

The dependence units are corpus-specific: the EverMemBench questions span five
shared topics, not 68 independent histories. Each source is used under a deterministic, disclosed audience overlay that keeps gold evidence authorized
for the reader; original evidence labels remain the recall ground truth. Thirty
LongMemEval unanswerable controls and four LoCoMo questions are answer-eligible but
have no positive-recall denominator (inapplicable, not zero). The full LongMemEval
frame forms one connected session component, so it is reported descriptively. All
2{,}218 GateMem mapping checkpoints were joined; none carries authoritative audience
and viewer labels, so no strict-audience cell is eligible. GroupMemBench has no
locally validated eligible frame and an unresolved redistribution permission. Current coverage is in Table~\ref{tab:coverage}.

\paragraph{Native frame.} \blore{} is evaluated on an outcome-blind frame of all
representable controlled scenarios: 42 of 43 eligible scenarios were selected
(one excluded before outcomes because a public anonymous viewer has no exact
native viewer state), each with five answers. The frame supports description of
the evaluated implementation only.

\paragraph{Conformance suites.} An executable reference oracle, independent of
every query implementation, derives expected admission and reason codes.
Randomized small universes of participants, items, audiences, viewers,
transformations and grants are compared against every write, derivation, grant,
retrieval, direct-lookup and context-assembly path of \bflat{} and \bgraph{}, and
against each inventoried native path of \blore{}. Mutation operators
(App.~\ref{app:mutations}) inject defects such as reversing the subset relation,
treating unknown viewers as owner-private, unioning derived audiences or
separately approved grants, admitting incomplete-provenance summaries, letting a
grant authorize a new derivation, and authorizing after ranking.

\subsection{Answer judging and pre-confirmation review}
\label{app:judging}

\paragraph{Metadata.} The judging routes and their amendment history are in App.~\ref{app:deviations} (item 7). Session identifiers and
exposed settings are retained privately; for the coding-assistant and \texttt{gpt-6-sol} routes, served model version,
per-judgment token usage and provider response identifiers are unavailable and
recorded as unknown; for Claude judgments the transport verified the served model
from the response stream and retained usage. Recorded timestamps measure
artifact creation, not model deliberation. Each judgment stores the rationale,
exact evidence quotations and uncertainty, and passes the frozen review schema and
citation validation; that validation binds what was judged but does not certify
that the judgment is correct.

\paragraph{Qualification.} For the first session, eighteen excluded, stipulated
cases (negation, supersession, attribution, invariance, ambiguity and admission
failure) all matched their expected labels; that session had seen the key, so
this was non-blinded consistency qualification, and one case was a stipulated
abstention rather than a newly generated response. Each later \texttt{gpt-6-sol} context was qualified individually on the same
unchanged controls; fresh-session Claude judges share one procedure-level
qualification reused unchanged across windows. No judge context saw keys, prior
labels or other assignments; a failed qualification is not retried until it agrees \unskip.

\paragraph{Scope and completion.} The review queue has 15{,}229 unique subjects:
15{,}193 original responses shared by both views, reviewed once, plus 36
recovered responses that need their own review.  All 15{,}229 subjects carry a validated, durably saved judgment and
none is pending: 170 parked subjects were admitted from existing saved judgments (86 held
exact-projection records previously validated; 84 saved responses validated by the
unchanged validator at admission), and 107 others
(malformed, exposure-invalid, uncertain or non-verbatim-citation cases, and one
held batch) received a single fresh judgment with no further holds. The judgments
mix routes: an initial coding-assistant session, parallel \texttt{gpt-6-sol}
sessions and fresh-session \texttt{claude-opus-5-5} judges, with the 84 admitted
saved responses split between the latter two; each route is a recorded provenance
stratum, and every record is marked as not independently validated. The last two
judging windows were operated and checked by the executing AI coding session under
study approval, neither independently nor by a human.{}

\paragraph{Audit.} The approved audit re-assesses up to 500 operationally negative answers per
target arm, drawn with a pre-recorded seed from frozen pools. Given evidence of rubric-verification
noise and weak judge sensitivity \citep{peng2026ruverbench,bagaria2026rubric,chen2026judge}, judging
and audit are AI consistency evidence, not independent validation or human ground truth. Pool
membership and the seed are frozen before any draw; sampled identities and
denominators are preserved, with no adaptive replacement. Any resulting bound is
conditional on the AI label procedure and does not certify semantic
accuracy. No audit sample has been drawn.

\paragraph{Pre-confirmation review.} Before confirmation, review used
study-approved
AI-assisted judgments plus ten direct judgments by a study author presented
without arm or system labels; the ten comparisons agree descriptively (10/10),
but repeat concealment was not established and no independent inter-rater
reliability was measured.

\subsection{Native system validation (full)}
\label{app:native-full}

All 42 selected native scenarios completed with five answers each (210 answers).
Under executing-session AI review, 195 of 210 answers were marked correct, with
15 incorrect abstentions and 10 correct abstentions; retrieved gold credit was
180 of 200 eligible answer-level gold items. These counts are descriptive: the
repeated answers are not 210 independent cases, the labels are AI reviews rather
than human judgments, and the cases span earlier revisions of the native code rather
than one homogeneous repaired candidate. Four unfavorable cases are retained:
approved owner-private widening (five incorrect abstentions; context from another conversation), complete intersection (five incorrect abstentions), overlapping
audiences (five incorrect abstentions; the requested group's evidence was ranked
second), and summary consolidation (correct answers without full consolidated
gold recall). Two coverage qualifications apply: the selected native program did
not execute an independent direct-identifier probe for existence opacity, and the
incomplete-provenance cases recorded non-emitting requests rather than runtime
refusals. 

\paragraph{Scoped native qualification.} A separately accepted qualification of
the selected native configuration passed all 39 phases (three controls, a
collection phase and 35 selected modules; 377 unique selected tests, 1{,}048.6
seconds, no model calls). The acceptance checked 316 declared trace-file hashes,
the exact collected and executed test union, boundary records and cleanup. This
supports the selected configuration and checks only: not arbitrary native
functionality, the reduced public artifact, or new semantic utility. Three
historical native evidence references named by the integration record remain
unauthenticated, and current success cannot authenticate them retrospectively.

\subsection{Supporting studies and negative results (full)}
\label{app:supporting-full}

\begin{table}[h]
\centering
\small
\caption{Required evidence tracks (exact joins). Reproduced is not scientifically
accepted, an output is not a graded answer, and a missing receipt is not a zero.}
\label{tab:coverage}
\begin{tabular}{@{}lll@{}}
\toprule
Track & Expected frame & Retained coverage \\
\midrule
Identity policies (Study A) & 28 cells + 4 controls & 32 executed; tables reproduced \\
Matched availability (H4b) & 16 cells & 16 executed; tables reproduced \\
Systems, original (Study B) & 600 samples & 600 retained; all 120 retrievals denied \\
Systems, companion & 840 timed calls & 840 executed; contended host \\
External retrieval & 2{,}549 questions & 2{,}549 prepared (15{,}294 cells) \\
External answers & 7{,}647 identities & 7{,}647 outputs; not graded \\
GateMem mapping & 2{,}218 checkpoints & 0 eligible (no audience labels) \\
Installation census & 14 metrics & None; 7 controlled workloads instead \\
Native qualification & 39 phases & 39 passed (377 tests); scoped \\
Native evidence references & 3 artifacts & Unauthenticated \\
\bottomrule
\end{tabular}
\end{table}

\paragraph{Status.} The identity-policy, matched-availability and systems
studies below were executed locally without model calls. An independent review reproduced all four retained tables
(Tables~\ref{tab:supporting}--\ref{tab:census}) from their unchanged inputs and
accepted their packaging; that is technical reproduction, not independent
scientific acceptance, so we report them as controlled, descriptive evidence
within the scopes stated below, and no headline claim depends on them.

\begin{table}[h]
\centering
\small
\caption{Supporting studies (tables reproduced; not independently accepted). Top: origin-level
probe opportunities across seven identity-fault families before reconciliation
(controls excluded); denominators count controlled probes, not participants or histories. Bottom: matched derivation cases over both reference architectures; availability
is direct lookup of the eligible derived item, reported separately from top-1
retrieval inclusion.
Generated from pinned exports.}
\label{tab:supporting}
\input{generated/supporting_identity_utility}
\end{table}

\textbf{Identity evidence (Study A).} Seven fault families (conflicting, extra,
forged, missing, stale, unknown and unresolved identity evidence) were crossed with
four declared policies, plus an exact one-to-one control through each
(Table~\ref{tab:supporting}). Before reconciliation, fail-open exposed forbidden
items in 11 of 13 opportunities; the three fail-closed alternatives exposed none
but lost entitled evidence at different rates. After attested reconciliation,
every origin and derived decision matched its known entitlement, all 16 superseded
copies were denied, and all four controls kept owner and authorized access while
denying other participants. This is a reference-policy experiment, not a change
to \blore{} or the reference stores; ``provisional'' is one declared conservative
operationalization (owner-only escrow until reconciliation).

\textbf{Derivation availability (H4b).} Across eight matched cases (four derivation
families on both architectures), the intersection item was available in all eight and ranked first in four;
blanket suppression made none available
(Table~\ref{tab:supporting}). These are case counts, not population rates.

\textbf{Systems realization (Study B).} Matched measurements cover two
architectures, two representations (audience labels and principal postings), five
operations and 30 repetitions (600 samples) on warm-cache SQLite. All 120 of their
retrieval samples returned empty results (denial), so they do not describe
successful retrieval latency. The companion studies add 840 timed owner calls (120 positive retrieval and 720
lifecycle and materialization; 30 per cell over 28 cells) after 140 excluded
warm-ups (Table~\ref{tab:systems}). Timings use warm SQLite and operating-system
caches on one laptop-class machine (10 cores, solid-state storage, SQLite
3.51.3), with unrelated background activity of about half a core recorded, so
they are descriptive, not uncontended or production latency, and none concerns
native latency. The timer covers the durable owner operation including its
authorization index; the source-missing operation measures owner-visible state,
not physical purge, and reopen with an idempotent write is not crash recovery.
We also count objects, audiences and grants in seven controlled workload
prestates, each realized in four physical configurations (Table~\ref{tab:census});
this is not an installation census, which needs a production deployment that does
not exist.

\begin{table}[h]
\centering
\small
\caption{Companion systems timings, nearest-rank median / 95th percentile in
milliseconds over 30 repetitions per cell (warm cache; contended host; generated
from the pinned table).}
\label{tab:systems}
\input{generated/supporting_systems}
\vspace{0.7em}
\caption{Controlled workload prestates used by the systems study (logical counts;
not deployment data).}
\label{tab:census}
\input{generated/supporting_census}
\end{table}

{}

\paragraph{Held-out pilot.} The corrected historical pilot report covers draws
8--39: 32 paired histories, 1{,}920 retrieval cells and 479 valid answer records
from 480 logical answers and 483 physical attempts, with one retained terminal
failure. Retained AI labels mark 160/160 \bflat{}, 157/159 \bgraph{} and
120/160 unscoped answers correct (unscoped: 25 wrong non-abstaining answers and
15 abstentions), with no wrong-principal labels. These are descriptive labels,
not verified correctness; the primary unit is the 32 paired histories, not 479
observations, and zero wrong-principal labels neither show the planned reduction
nor prove a population zero. The report binds its inputs but leaves the exact
executed producer revision of the older pilot reports unresolved.

\begin{figure}[h]
\centering
\input{generated/topk_figure}
\caption{Mean entitled Recall@$k$ over 10{,}000 histories (six questions, $d=8$, fetch budget 20);
the dotted line marks the frozen $k=5$.}
\label{fig:topk}
\end{figure}
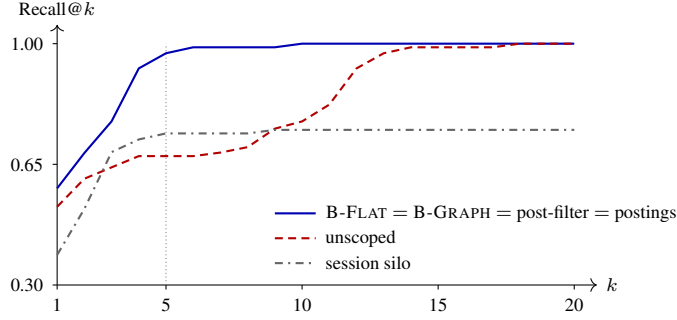

\paragraph{Retrieval budget.} Every confirmation cell stores its 20 ranked
candidates, so Recall@$k$ follows for $k=1,\dots,20$ at the frozen fetch budget without new runs
(recomputed $k=5$ equals the stored score in all 360{,}000 dose-8 cells). At every $k$ the
architectures equal the policy-equivalent post-filter and postings exactly and are never below unscoped or
silo retrieval; they reach full recall at $k=\TopkArchFull$, unscoped retrieval only at
$k=\TopkUnscFull$, and the silo plateaus at \TopkSiloMax{} (Figure~\ref{fig:topk}).

\paragraph{Negative, null and weak results.}

We retain these results as reported:{} the frozen H3a/b test could not
detect a substitution reduction at the observed base rate, so the joint decision is
not established (\S\ref{sec:results-wrong-principal});{} the held-out pilot of 32 histories did not demonstrate a wrong-principal substitution reduction, so the planned effect sizes
remained planning targets; four unfavorable native cases (three with incorrect abstentions, one with incomplete gold recall; \S\ref{sec:results-native}); 85 original terminal citation-contract failures, 18
of which remain unresolved under recovery; and, in a 75-answer supporting slice
(39 controlled and 36 external answers; descriptive AI review), 47 correct
answers, 27 valid misses and one terminal failure, where twelve LoCoMo misses
concern questions that lack upstream evidence annotations.
\unskip

%% file: generated/external_strata.tex
\setlength{\tabcolsep}{4pt}\begin{tabular}{@{}lrrrrr@{}}
\toprule
Corpus & Questions & \texttt{gpt-5.6-luna} & \texttt{gpt-6-luna} & \texttt{claude-opus-5-5} & Mixed-model q. \\
\midrule
LongMemEval & 500 & 0 & 86 & 1{,}414 & 81 \\
LoCoMo & 1{,}981 & 1{,}807 & 3{,}238 & 898 & 46 \\
EverMemBench & 68 & 0 & 15 & 189 & 13 \\
\midrule
Total & 2{,}549 & 1{,}807 & 3{,}339 & 2{,}501 & 140 \\
\bottomrule
\end{tabular}

%% file: generated/supporting_identity_utility.tex
\begin{tabular}{@{}lcc@{}}
\toprule
Identity policy (before reconciliation) & Forbidden exposed & Entitled lost \\
\midrule
Fail open & 11/13 & 1/43 \\
Public-only fallback & 0/13 & 15/43 \\
Quarantine (delayed binding) & 0/13 & 27/43 \\
Provisional, later reconciled & 0/13 & 11/43 \\
\midrule
Derivation treatment (8 cases) & Derived item available & Ranked first \\
\midrule
Intersection & 8/8 & 4/8 \\
Blanket suppression & 0/8 & 0/8 \\
\bottomrule
\end{tabular}

%% file: generated/supporting_systems.tex
\begin{tabular}{@{}lcccc@{}}
\toprule
& \multicolumn{2}{c}{\bflat{}} & \multicolumn{2}{c}{\bgraph{}} \\
\cmidrule(lr){2-3}\cmidrule(lr){4-5}
Operation (ms, p50 / p95) & Audience labels & Postings & Audience labels & Postings \\
\midrule
Positive retrieval & 575 / 597 & 585 / 604 & 815 / 855 & 812 / 837 \\
Derived materialization & 12.0 / 12.5 & 11.6 / 12.1 & 12.1 / 12.4 & 11.9 / 12.8 \\
Grant materialization & 10.6 / 11.3 & 10.5 / 11.3 & 10.6 / 11.3 & 10.2 / 10.6 \\
Grant removal & 0.56 / 0.87 & 0.68 / 1.48 & 0.50 / 0.60 & 0.46 / 0.59 \\
Artifact retirement & 0.71 / 0.98 & 0.65 / 0.84 & 0.55 / 0.70 & 0.45 / 0.55 \\
Source-missing state & 0.60 / 0.83 & 0.53 / 0.88 & 0.52 / 0.69 & 0.47 / 0.61 \\
Reopen + idempotent replay & 11.6 / 12.3 & 11.7 / 12.6 & 11.8 / 12.2 & 11.4 / 12.0 \\
\bottomrule
\end{tabular}

%% file: generated/supporting_census.tex
\begin{tabular}{@{}lrrrr@{}}
\toprule
Controlled workload & Objects & Sources & Audiences & Grants \\
\midrule
Positive retrieval & 134 & 134 & 2 & 0 \\
Derived materialization & 66 & 66 & 3 & 0 \\
Grant materialization & 66 & 66 & 3 & 0 \\
Artifact retirement & 67 & 66 & 3 & 0 \\
Source-missing state & 67 & 66 & 3 & 0 \\
Grant removal & 66 & 68 & 3 & 1 \\
Reopen + idempotent replay & 67 & 66 & 3 & 0 \\
\bottomrule
\end{tabular}

%% file: generated/topk_figure.tex
\includegraphics[width=257.33461pt,height=123.34775pt]{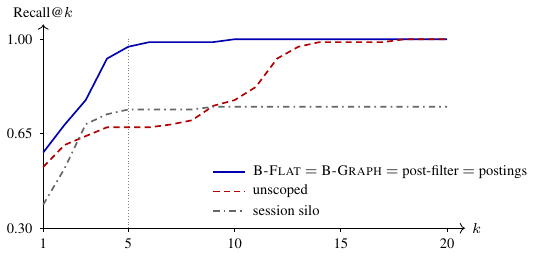}

%% file: references.bib
@inproceedings{afr2026,
  title = {Authorization-First Retrieval: Enforcing Least Privilege in Multi-Agent {RAG} Systems},
  author = {Namboothiri, Rohith},
  booktitle = {Proceedings of the 6th Workshop on Trustworthy NLP (TrustNLP 2026)},
  pages = {256--271},
  year = {2026},
  publisher = {Association for Computational Linguistics},
  doi = {10.18653/v1/2026.trustnlp-main.15},
  url = {https://aclanthology.org/2026.trustnlp-main.15/}
}

@misc{rezazadeh2025collaborativememory,
  title = {Collaborative Memory: Multi-User Memory Sharing in {LLM} Agents with Dynamic Access Control},
  author = {Rezazadeh, Alireza and Li, Zichao and Lou, Ange and Zhao, Yuying and Wei, Wei and Bao, Yujia},
  year = {2025},
  eprint = {2505.18279},
  archiveprefix = {arXiv},
  primaryclass = {cs.MA},
  url = {https://arxiv.org/abs/2505.18279}
}

@misc{gatemem2026,
  title = {{GateMem}: Benchmarking Memory Governance in Multi-Principal Shared-Memory Agents},
  author = {Ren, Zhe and Yang, Yibo and Chen, Yimeng and Zhao, Zijun and Fu, Benshuo and Shu, Zhihao and Zhang, Bingjie and Xu, Yangyang and Guo, Dandan and Yan, Shuicheng},
  year = {2026},
  eprint = {2606.18829},
  archiveprefix = {arXiv},
  primaryclass = {cs.LG},
  url = {https://arxiv.org/abs/2606.18829}
}

@misc{pisas2026,
  title = {{PiSAs}: Benchmarking Contextual Integrity in Multi-User Agentic Systems},
  author = {Gupta, Shubham and Mohammadi Sepahvand, Nazanin and Kumar, Abhinav and Subakan, Cem and Gella, Spandana and No{\"e}l, Pierre-Andr{\'e} and Taslakian, Perouz and Bagdasarian, Eugene and Zantedeschi, Valentina},
  year = {2026},
  eprint = {2607.05318},
  archiveprefix = {arXiv},
  primaryclass = {cs.MA},
  url = {https://arxiv.org/abs/2607.05318}
}

@misc{authmembench2026,
  title = {When Memory Becomes Authority: Benchmarking Authority Collapse at the Memory Consolidation Boundary},
  author = {Zhan, Qiuyang and Zhang, Rui and Guo, Sheng and Zhao, Lepeng and Liu, Zhuotao},
  year = {2026},
  eprint = {2608.01679},
  archiveprefix = {arXiv},
  primaryclass = {cs.AI},
  url = {https://arxiv.org/abs/2608.01679}
}

@misc{tmanm2026,
  title = {Securing {LLM}-Agent Long-Term Memory Against Poisoning: Non-Malleable, Origin-Bound Authority with Machine-Checked Guarantees},
  author = {Louck, Yedidel},
  year = {2026},
  eprint = {2606.24322},
  archiveprefix = {arXiv},
  primaryclass = {cs.CR},
  url = {https://arxiv.org/abs/2606.24322}
}

@misc{ppmf2026,
  title = {Memory Provenance Laundering in {LLM} Agents: A Non-Amplification Firewall for Persistent Memory},
  author = {Xu, Jinghan and Xiao, Yiyong and Shao, Wanru and Liu, Hankai and Li, Xinjin},
  year = {2026},
  eprint = {2607.29167},
  archiveprefix = {arXiv},
  primaryclass = {cs.CR},
  url = {https://arxiv.org/abs/2607.29167}
}

@misc{mapgraph2026,
  title = {{MAP-Graph}: Provenance-Aware Shared Memory for Multi-Agent Workflows},
  author = {Wang, Yiqi and Yan, Zihao and Zhang, Jiaqi and Wu, Zhangkai and Zheng, Mingkai and Sun, Zequn and Zhu, Yanming and Cai, Taotao},
  year = {2026},
  eprint = {2608.10509},
  archiveprefix = {arXiv},
  primaryclass = {cs.AI},
  url = {https://arxiv.org/abs/2608.10509}
}

@misc{memsecbench2026,
  title = {{MemSecBench}: Tracking Agent Memory Poisoning from Persistence to Consequence and Repair},
  author = {Chen, Xuanze and Xie, Xukang and Fu, Wentao and Zhou, Jiajun and Yu, Shanqing and Xuan, Qi},
  year = {2026},
  eprint = {2607.27080},
  archiveprefix = {arXiv},
  primaryclass = {cs.CR},
  url = {https://arxiv.org/abs/2607.27080}
}

@misc{mutmem2026,
  title = {{MutMem}: Cryptographically Authorized Mutation in Persistent Agent Memory},
  author = {Saidi, Walid},
  year = {2026},
  eprint = {2608.02843},
  archiveprefix = {arXiv},
  primaryclass = {cs.CR},
  url = {https://arxiv.org/abs/2608.02843}
}

@misc{superlocalmemory2026,
  title = {{SuperLocalMemory} 4.0: The Governed Memory Operating System for {AI} Agents},
  author = {Bhardwaj, Varun Pratap and Singh, Garima and Bhardwaj, Arun Pratap},
  year = {2026},
  eprint = {2608.08253},
  archiveprefix = {arXiv},
  primaryclass = {cs.AI},
  note = {Version 2},
  url = {https://arxiv.org/abs/2608.08253}
}

@inproceedings{mireshghallah2025cimemories,
  title = {{CIMemories}: A Compositional Benchmark For Contextual Integrity In {LLMs}},
  author = {Mireshghallah, Niloofar and Mangaokar, Neal and Kokhlikyan, Narine and Zharmagambetov, Arman and Zaheer, Manzil and Mahloujifar, Saeed and Chaudhuri, Kamalika},
  booktitle = {International Conference on Learning Representations (ICLR)},
  year = {2026},
  eprint = {2511.14937},
  archiveprefix = {arXiv},
  url = {https://proceedings.iclr.cc/paper_files/paper/2026/hash/9a2bcfaf383638e166162a25b6dff125-Abstract-Conference.html}
}

@misc{eal2026,
  title = {Agent Memory Is a Surface for Endogenous Authorization Laundering},
  author = {Cerruti, Tommaso and Okamoto, Mika and Erol, Ansel Kaplan},
  year = {2026},
  eprint = {2609.01836},
  archiveprefix = {arXiv},
  primaryclass = {cs.CR},
  url = {https://arxiv.org/abs/2609.01836}
}

@misc{memclaw2026,
  title = {Governed Shared Memory for Multi-Agent {LLM} Systems},
  author = {Margalit, Yanki and Cohen-Inger, Nurit and Avram, Erni and Taig, Ran and Margalit, Oded},
  year = {2026},
  eprint = {2606.24535},
  archiveprefix = {arXiv},
  primaryclass = {cs.AI},
  url = {https://arxiv.org/abs/2606.24535}
}

@misc{chen2026deploymentmemorization,
  title = {Deployment-Time Memorization in Foundation-Model Agents},
  author = {Chen, Lei {(Rachel)} and Zhang, Guilin and Zhao, Kai and Cirne, Dalmo and Olsen, Andy and Chu, Xu and Miller, Zeke and Blanken, Alet and Anoun, Amine and Ting, Jerry},
  year = {2026},
  eprint = {2606.10062},
  archiveprefix = {arXiv},
  primaryclass = {cs.AI},
  note = {ICML 2026 Workshop on Memory in Foundation Models (MemFM), per arXiv comments},
  url = {https://arxiv.org/abs/2606.10062}
}

@misc{liu2026authorization,
  title = {Authorization Before Context: A Model-Neutral Audience Boundary Against Cross-Audience Memory Leakage in Agentic Systems},
  author = {Liu, Sibo},
  year = {2026},
  eprint = {2608.17148},
  archiveprefix = {arXiv},
  primaryclass = {cs.CR},
  note = {Non-archival workshop paper, AdvML-Frontiers x CoTMA at COLM 2026},
  url = {https://arxiv.org/abs/2608.17148}
}

@misc{yang2026groupmembench,
  title = {{GroupMemBench}: Benchmarking {LLM} Agent Memory in Multi-Party Conversations},
  author = {Yang, Jingbo and Lai, Kwei-Herng and Wang, Xiaowen and Chang, Shiyu and Harari, Yaar and Gabrilovich, Evgeniy},
  year = {2026},
  eprint = {2605.14498},
  archiveprefix = {arXiv},
  primaryclass = {cs.CL},
  url = {https://arxiv.org/abs/2605.14498}
}

@misc{evermembench2026,
  title = {Evaluating Long-Horizon Memory for Multi-Party Collaborative Dialogues},
  author = {Hu, Chuanrui and Li, Tong and Gao, Xingze and Chen, Hongda and Bai, Yi and Xu, Dannong and Lin, Tianwei and Li, Xiaohong and Han, Yunyun and Pei, Jian and Deng, Yafeng},
  year = {2026},
  eprint = {2602.01313},
  archiveprefix = {arXiv},
  primaryclass = {cs.CL},
  url = {https://arxiv.org/abs/2602.01313}
}

@misc{yang2026multiuser,
  title = {Multi-User Large Language Model Agents},
  author = {Yang, Shu and Zhu, Shenzhe and Zhu, Hao and Enr{\'i}quez, Jos{\'e} Ram{\'o}n and Wang, Di and Pentland, Alex and Bakker, Michiel A. and Pei, Jiaxin},
  year = {2026},
  eprint = {2604.08567},
  archiveprefix = {arXiv},
  primaryclass = {cs.CL},
  url = {https://arxiv.org/abs/2604.08567}
}

@inproceedings{wu2025longmemeval,
  title = {{LongMemEval}: Benchmarking Chat Assistants on Long-Term Interactive Memory},
  author = {Wu, Di and Wang, Hongwei and Yu, Wenhao and Zhang, Yuwei and Chang, Kai-Wei and Yu, Dong},
  booktitle = {International Conference on Learning Representations (ICLR)},
  year = {2025},
  url = {https://proceedings.iclr.cc/paper_files/paper/2025/hash/d813d324dbf0598bbdc9c8e79740ed01-Abstract-Conference.html}
}

@inproceedings{maharana2024locomo,
  title = {Evaluating Very Long-Term Conversational Memory of {LLM} Agents},
  author = {Maharana, Adyasha and Lee, Dong-Ho and Tulyakov, Sergey and Bansal, Mohit and Barbieri, Francesco and Fang, Yuwei},
  booktitle = {Proceedings of the 62nd Annual Meeting of the Association for Computational Linguistics (Volume 1: Long Papers)},
  pages = {13851--13870},
  year = {2024},
  publisher = {Association for Computational Linguistics},
  doi = {10.18653/v1/2024.acl-long.747},
  url = {https://aclanthology.org/2024.acl-long.747/}
}

@misc{debenedetti2025camel,
  title = {Defeating Prompt Injections by Design},
  author = {Debenedetti, Edoardo and Shumailov, Ilia and Fan, Tianqi and Hayes, Jamie and Carlini, Nicholas and Fabian, Daniel and Kern, Christoph and Shi, Chongyang and Terzis, Andreas and Tram{\`e}r, Florian},
  year = {2025},
  eprint = {2503.18813},
  archiveprefix = {arXiv},
  primaryclass = {cs.CR},
  url = {https://arxiv.org/abs/2503.18813}
}

@misc{costa2025fides,
  title = {Securing {AI} Agents with Information-Flow Control},
  author = {Costa, Manuel and K{\"o}pf, Boris and Kolluri, Aashish and Paverd, Andrew and Russinovich, Mark and Salem, Ahmed and Tople, Shruti and Wutschitz, Lukas and Zanella-B{\'e}guelin, Santiago},
  year = {2025},
  eprint = {2505.23643},
  archiveprefix = {arXiv},
  primaryclass = {cs.CR},
  url = {https://arxiv.org/abs/2505.23643}
}

@misc{palumbo2026forge,
  title = {Formal Policy Enforcement for Real-World Agentic Systems},
  author = {Palumbo, Nils and Choudhary, Sarthak and Choi, Jihye and Amir, Guy and Chalasani, Prasad and Jha, Somesh},
  year = {2026},
  eprint = {2602.16708},
  archiveprefix = {arXiv},
  primaryclass = {cs.CR},
  url = {https://arxiv.org/abs/2602.16708}
}

@inproceedings{pang2019zanzibar,
  title = {Zanzibar: {Google's} Consistent, Global Authorization System},
  author = {Pang, Ruoming and Caceres, Ramon and Burrows, Mike and Chen, Zhifeng and Dave, Pratik and Germer, Nathan and Golynski, Alexander and Graney, Kevin and Kang, Nina and Kissner, Lea and Korn, Jeffrey L. and Parmar, Abhishek and Richards, Christina D. and Wang, Mengzhi},
  booktitle = {2019 USENIX Annual Technical Conference (USENIX ATC 19)},
  year = {2019},
  pages = {33--46},
  publisher = {USENIX Association},
  url = {https://www.usenix.org/conference/atc19/presentation/pang}
}

@techreport{bell1976secure,
  title = {Secure Computer System: Unified Exposition and {Multics} Interpretation},
  author = {Bell, D. Elliott and La Padula, Leonard J.},
  institution = {MITRE Corporation},
  address = {Bedford, MA},
  year = {1976},
  month = mar,
  note = {DTIC accession ADA023588},
  doi = {10.21236/ADA023588},
  url = {https://doi.org/10.21236/ADA023588}
}

@article{denning1976lattice,
  title = {A Lattice Model of Secure Information Flow},
  author = {Denning, Dorothy E.},
  journal = {Communications of the ACM},
  volume = {19},
  number = {5},
  pages = {236--243},
  year = {1976},
  month = may,
  doi = {10.1145/360051.360056},
  url = {https://doi.org/10.1145/360051.360056}
}

@inproceedings{myers1997decentralized,
  title = {A Decentralized Model for Information Flow Control},
  author = {Myers, Andrew C. and Liskov, Barbara},
  booktitle = {Proceedings of the Sixteenth ACM Symposium on Operating Systems Principles (SOSP)},
  pages = {129--142},
  year = {1997},
  month = oct,
  publisher = {ACM},
  doi = {10.1145/268998.266669},
  url = {https://doi.org/10.1145/268998.266669}
}

@article{sabelfeld2009declassification,
  title = {Declassification: Dimensions and Principles},
  author = {Sabelfeld, Andrei and Sands, David},
  journal = {Journal of Computer Security},
  volume = {17},
  number = {5},
  pages = {517--548},
  year = {2009},
  doi = {10.3233/JCS-2009-0352},
  url = {https://doi.org/10.3233/JCS-2009-0352}
}

@inproceedings{bates2013provenance,
  title = {Towards Secure Provenance-Based Access Control in Cloud Environments},
  author = {Bates, Adam and Mood, Ben and Valafar, Masoud and Butler, Kevin},
  booktitle = {Proceedings of the Third ACM Conference on Data and Application Security and Privacy (CODASPY '13)},
  pages = {277--284},
  year = {2013},
  publisher = {ACM},
  doi = {10.1145/2435349.2435389},
  url = {https://sts.cs.illinois.edu/papers/paper/TowardsSecureProvena20130219.html}
}

@inproceedings{denhartog2016datafusion,
  title = {A Policy Framework for Data Fusion and Derived Data Control},
  author = {den Hartog, Jerry and Zannone, Nicola},
  booktitle = {Proceedings of the 2016 ACM International Workshop on Attribute Based Access Control (ABAC '16)},
  pages = {47--57},
  year = {2016},
  publisher = {ACM},
  doi = {10.1145/2875491.2875492},
  url = {https://doi.org/10.1145/2875491.2875492}
}

@book{nissenbaum2009privacy,
  title = {Privacy in Context: Technology, Policy, and the Integrity of Social Life},
  author = {Nissenbaum, Helen},
  publisher = {Stanford University Press},
  year = {2009},
  doi = {10.1515/9780804772891},
  url = {https://www.sup.org/books/law/privacy-context}
}

@article{chronis2025filtered,
  title = {Filtered Vector Search: State-of-the-Art and Research Opportunities},
  author = {Chronis, Yannis and Caminal, Helena and Papakonstantinou, Yannis and {\"O}zcan, Fatma and Ailamaki, Anastasia},
  journal = {Proceedings of the VLDB Endowment},
  volume = {18},
  number = {12},
  pages = {5488--5492},
  year = {2025},
  doi = {10.14778/3750601.3750700},
  url = {https://www.vldb.org/pvldb/vol18/p5488-caminal.pdf}
}

@inproceedings{maurer2009empirical,
  title = {Empirical {Bernstein} Bounds and Sample Variance Penalization},
  author = {Maurer, Andreas and Pontil, Massimiliano},
  booktitle = {Conference on Learning Theory (COLT)},
  year = {2009},
  url = {https://www.cs.mcgill.ca/~colt2009/papers/012.pdf}
}

@misc{chen2026judge,
  title         = {A Judge Should Know What Changed: Construct Validity for {LLM}-as-a-Judge Evaluation},
  author        = {Chen, Jianlin and Chen, Wenhui and Lin, Ziyao and Vong, Chi Man},
  year          = {2026},
  eprint        = {2608.24419},
  archivePrefix = {arXiv},
  url           = {https://arxiv.org/abs/2608.24419}
}

@misc{peng2026ruverbench,
  title         = {Can {LLM}-as-a-Judge Reliably Verify Rubrics in Agentic Scenarios?},
  author        = {Peng, Yangda and Qi, Yunjia and Xia, Haotian and He, Guanzhong and Shi, Xintong and Xuan, Richeng and Lu, Songyuanyi and Liu, Yixian and Hu, Zhichao and Liu, Yuhong and Peng, Hao},
  year          = {2026},
  eprint        = {2606.29920},
  archivePrefix = {arXiv},
  note          = {Version 2},
  url           = {https://arxiv.org/abs/2606.29920}
}

@misc{bagaria2026rubric,
  title         = {Judging {LLM}-as-a-Judge: Concerning Rubric Artifacts in {LLM}-based Automated Text Generation Evaluation},
  author        = {Bagaria, Anshul and Sundaram, Sowmya S and Krishnan, Gokul S and Ravindran, Balaraman},
  year          = {2026},
  eprint        = {2609.02942},
  archivePrefix = {arXiv},
  url           = {https://arxiv.org/abs/2609.02942}
}
